\documentclass[preprint]{aastex}

\makeatletter
\let\@dates\relax
\makeatother

\usepackage{amsmath}
\usepackage{color}
\usepackage{graphicx}
\usepackage{tikz}
\usetikzlibrary{calc,arrows}

\shorttitle{Hydrosimulations of a compact source for G2}

\begin{document}

\renewcommand{\thefootnote}{\alph{footnote}}

\title{Hydrodynamical simulations of a compact source scenario for the Galactic Center cloud G2} 
\author{A. Ballone\altaffilmark{1,2}, M. Schartmann\altaffilmark{1,2}, A. Burkert\altaffilmark{1,2,3}, S. Gillessen\altaffilmark{2}, R. Genzel\altaffilmark{2}, T.K. Fritz\altaffilmark{2}, F. Eisenhauer\altaffilmark{2}, O. Pfuhl\altaffilmark{2}, T. Ott\altaffilmark{2}} 

\altaffiltext{1}{University Observatory Munich, Scheinerstra{\ss}e 1, D-81679 M{\"u}nchen, Germany; aballone@mpe.mpg.de}
\altaffiltext{2}{Max-Planck-Institute for Extraterrestrial Physics, Postfach 1312, Giessenbachstra{\ss}e, D-85741 Garching, Germany}
\altaffiltext{3}{Max-Planck Fellow}

\begin{abstract}
The origin of the dense gas cloud G2 discovered in the Galactic Center \citep{Gillessen_12} is still a debated puzzle. G2 might be a diffuse cloud or the result of an outflow from an invisible star embedded in it. We present hydrodynamical simulations of the evolution of different spherically symmetric winds of a stellar object embedded in G2. We find that the interaction with the ambient medium and with the extreme gravitational field of the supermassive black hole in the Galactic Center must be taken into account for such a source scenario. The thermal pressure of the hot and dense atmosphere confines the wind, while its ram pressure shapes it via stripping along the orbit, with the details depending on the wind parameters. Tidal forces squeeze the wind near pericenter, reducing it to a thin and elongated filament. We also find that in this scenario most of the Br$\gamma$ luminosity is expected to come from the densest part of the wind, which has a highly filamentary structure with low filling factor.
For our assumed atmosphere, the observations can be best matched by a mass outflow rate of $\dot{M}\mathrm{_w=8.8\times 10^{-8} M_{\odot} \;yr^{-1}}$ and a wind velocity of $v\mathrm{_w = 50 \;km/s}$. These values are compatible with those of a young T Tauri star wind, as already suggested by \citet{Scoville_13}.

\end{abstract}


\section{Introduction}

Our Galactic Center hosts a supermassive black hole (SMBH) of $M\mathrm{_{BH}  \simeq 4.31 \times 10^6 \; M_{\odot}}$ \citep{Ghez_08, Gillessen_09} and a surrounding atmosphere of X-ray emitting hot gas \citep[e.g.][]{Baganoff_03,Yuan_03,Xu_06}. Therefore it is one of the most extreme and peculiar places in the Milky Way and a very interesting laboratory for astrophysics. In the last few months, the discovery of the dense gas clump G2 \citep{Gillessen_12,Gillessen_13} has caught the attention of the astronomical community. Observations in the L-band with the infrared imager NACO\footnote{http://www.eso.org/sci/facilities/paranal/instruments/naco/} at the Very Large Telescope \citep{Lenzen_98,Rousset_98} have revealed a warm dust component, with a temperature of $T\mathrm{_{dust}\approx550 \; K}$, while the integral field spectrograph SINFONI\footnote{http://www.eso.org/sci/facilities/paranal/instruments/sinfoni/} \citep{Eisenhauer_03, Bonnet_04} allowed the detection of Br$\gamma$, He I and Paschen-$\alpha$ line emission from an ionized gas ($T\mathrm{_{gas}\approx10^4 \; K}$) component, with a constant Br$\gamma$ luminosity $L\mathrm{_{Br\gamma} \approx 2\times 10^{-3} L_{\odot}}$ from 2004 to 2012. Assuming a homogeneous sphere of radius $R\mathrm{_{c} \approx 1.9\times10^{15} \; cm}$, \citet{Gillessen_12} derived, from the observed Br$\gamma$ luminosity, a density of approximately $\mathrm{\rho_c \approx 6.1\times10^{-19}\; g\; cm^{-3}}$, with a corresponding mass of $M\mathrm{_{G2} \approx 1.7 \times 10^{28}\; g \approx 3}$ Earth masses. More recent observations have also confirmed the presence of an extended “tail” (named G2t) with a Br$\gamma$ luminosity comparable to that of G2 and an estimated mass $M\mathrm{_{G2t} = 1 - 2} \;M\mathrm{_{G2}}$ \citep{Gillessen_13}.

With the help of observations from the last 10 yr, \citet{Gillessen_12, Gillessen_13} derived the dynamical properties of the object, finding that G2 and G2t are moving toward the SMBH of the Milky Way. They seem to lie on a common, very eccentric orbit ($e\mathrm{=0.966}$) with pericenter at only 2200 Schwarzschild radii, which G2 is expected to reach in late 2013. The strong tidal field has already produced a large velocity gradient in G2 ($\mathrm{230\; km\;s^{-1}}$ in 2008, $\mathrm{370\; km\;s^{-1}}$ in 2011, and $\mathrm{600 \;km\;s^{-1}}$ in 2012) and hydrodynamic effects due to the interaction with the hot and dense environment (i.e., ram pressure and hydro-instabilities) are also expected to play a role in the current dynamical evolution \citep{Burkert_12,Schartmann_12}.
Recent observations with the Keck telescope have also been published by \citet{Phifer_13}. With the help of the near-infrared camera NIRC2 \footnote{http://www2.keck.hawaii.edu/inst/nirc2/} and the OSIRIS\footnote{http://www2.keck.hawaii.edu/inst/osiris/} integral field spectrograph, these authors confirmed the detection of G2 with properties comparable to those found by \citet{Gillessen_12, Gillessen_13}, but with a small positional offset between the L-band and Br$\gamma$ emission and slightly different orbital parameters (eccentricity of $e \simeq 0.98$ and pericenter distance of 1900 Schwarzschild radii). Given the somewhat different orbit, these authors predict that the center of mass of G2 will have its closest approach to SgrA* in 2014 March. Of course, G2 is an extended object and it is expected to become strongly elongated due to tidal forces, so thinking of a single pericenter date is misleading. Indeed, new observations presented by \citet{Gillessen_13b} already show weak emission from material that has already passed pericenter and the duration of the pericenter passage has been estimated to be roughly 1 yr.
G2 offers us the possibility to study its fate ``on the fly'' and to eventually observe a minor accretion event on SgrA* in the next decades. Several new observations of the Galactic Center, in different bands of the electromagnetic spectrum, are already planned for the time of the pericenter passage\footnote{https://wiki.mpe.mpg.de/gascloud/FrontPage}.

A big challenge is to explain the origin and nature of such a dense and low angular momentum object.
Interestingly, \citet{Gillessen_12, Gillessen_13} find that the observed orbit of the G2 complex (G2+G2t) roughly lies in the plane of the clockwise rotating disk of young and massive stars ranging from $\mathrm{0.04 \; pc}$ to $\mathrm{0.5 \; pc}$ around the central hot bubble \citep{Genzel_03,Paumard_06,Bartko_09,Alig_11, Alig_13}. Even if this finding is still debated \citep{Phifer_13}, several authors have argued that the origin of G2 is related to the clockwise disk. One promising scenario would be a compact gas cloud that formed as a result of stellar wind interactions \citep{Cuadra_05,Cuadra_06,Gillessen_12,Burkert_12} on a highly eccentric orbit. \citet{Burkert_12}, \citet{Schartmann_12} and \citet{Anninos_12} have studied in high detail the evolution and fate of such an object with properties similar to those of G2.

Together with the diffuse gas cloud scenario, Gillessen et al. (2012) also suggested another interesting option, i.e., G2 could be the outflow from a low luminosity star that is too faint to be observed. \citet{Murray-Clay_12} have shown that multiple scatterings in the young stellar ring can put a low-mass star on a very high eccentricity orbit every $\mathrm{10^6\; yr}$ and the probability for a young star with an undisrupted protoplanetary disk to reach G2's pericenter for the first time is $\mathrm{\sim 0.1 \%}$. These authors show, with the help of analytical considerations, that the observed properties of G2 can be explained by gas outflowing from a photoevaporating protoplanetary disk (due to the high flux of far-ultraviolet and Lyman photons in the Galactic Center) and being tidally stripped while reaching SgrA*. \citet{Miralda-Escude_12} suggested a similar scenario, proposing that a close encounter with a stellar black hole could deflect an old low-mass star on a high-eccentricity orbit and during the same encounter produce a disk around it by stripping its outer envelope.
\citet{Meyer_12} investigated the possibility that a nova, on a similar orbit, could produce an expanding shell, following the Spherical Shell Scenario proposed and studied by \citet{Burkert_12} and \citet{Schartmann_12}.
Finally, \citet{Scoville_13} suggested that the observed emission could come from the tip of an inner, thin and cold bow shock, produced by the wind of a T Tauri star plunging into SgrA*. In their model, the Br$\gamma$ luminosity arises due to shell material, which is collisionally ionized by the wind. This model is attractive since the luminosity is given by the wind properties and hence remains constant with time over the observed $\mathrm{\sim 10\; yr}$.

Models involving a central source  have been studied up to now only with simplified analytical approximations that do not fully take into account the role of the hydrodynamic interaction with the ambient medium and of the SMBH tidal field in affecting the global structure of the outflows. The aim of our work is to study these effects for a large range of outflow parameters with the help of hydrodynamical simulations with the Eulerian code PLUTO \citep{Mignone_07,Mignone_12}, version 3.1.1.

In Section \ref{physwinds}, we give a brief overview of the physics of winds in the Galactic Center, discussing their structure and evolution with the help of analytical considerations. In Section \ref{setup}, we introduce our physical and numerical setup and our simulations are presented in Section \ref{results}. We give a global discussion and some remarks in Section \ref{discus} and our final summary and conclusions appear in Section \ref{sumconc}. In this work, we assume a Galactic center distance of $8.33 \;\mathrm{kpc}$ \citep{Gillessen_09}, so $1'' \simeq 1.25 \times 10^{17}\; \mathrm{cm} \simeq 0.04 \;\mathrm{pc}$.

\section{Physics of winds in the Galactic Center}\label{physwinds}

In this section, we provide some qualitative and quantitative considerations that will be helpful in understanding the structure and evolution of our simulated outflows. 

When spherical stellar winds interact with typical interstellar medium (ISM) gas ($n\mathrm{_{ISM}\approx 1 \;cm^{-3}}$, $P\mathrm{_{ISM}}\approx 10^{-12}\mathrm{ \;dyn \;cm^{-2}}$) they are well described by the model of \citet{Weaver_77}. Their structure consists of a \textit{free-wind region}, where the wind mass loss rate $\dot{M}\mathrm{_w}$ and the wind velocity $v\mathrm{_w}$ are constant with radius $r$, so that the gas density

\begin{equation}\label{densform}
\begin{split}
&\rho_w(\mathit{r})= \frac{\dot{M}_w}{4\pi v_w r^2}\approx\\
&\approx 4.47\times 10^{-16} \left(\frac{\dot{M}_w}{10^{-7} \mathrm{M_{\odot} /yr}}\right)\left(\frac{v_w}{50 \mathrm{\;km/s}}\right)^{-1}\left(\frac{r}{1\mathrm{\; AU}}\right)^{-2} \mathrm{ g\;cm^{-3}} 
\end{split}
\end{equation}

scales with $1/\mathit{r}^2$.
An inner shock is separating the free-wind region from a large and almost isobaric region of shocked stellar wind. A contact discontinuity separates the wind material from a dense shell of swept-up ambient medium whose outer boundary is an external shock propagating into the unperturbed ISM. After a phase of adiabatic expansion, the thickness of this outer shell reduces and the density increases substantially because radiative cooling occurs in this region on a relatively short timescale.

In our work, the wind interacts with the very hot and dense atmosphere present in the Galactic Center \citep[e.g.,][]{Baganoff_03,Yuan_03,Xu_06}. This atmosphere has extremely high thermal pressures, ranging from roughly $\mathrm{10^{-7} dyn \;cm^{-2}}$ at $0.05 \;\mathrm{pc}$ to roughly $ \mathrm{5\times10^{-4} dyn \;cm^{-2}}$ at around $1000$ Schwarzschild radii from SgrA* and a sound speed $c\mathrm{_s}$ ranging from $\approx 500 \; \mathrm{km \;s^{-1}}$ to $\approx 3000 \; \mathrm{km\;s^{-1}}$. Furthermore, in all the proposed scenarios for G2's origin related to outflows, given G2's small size and mass, the suggested mass-loss rates and velocities are always typical of relatively weak winds (weak with respect to their power, which is given by $L\mathrm{_w} = 0.5\dot{M}_\mathrm{w}v_\mathrm{w}^2$). As a result, the structure of winds in the proximity of Sgr A* is different from that of \citet{Weaver_77}. As in this classic model, our winds have a free-wind region, with an inner shock separating this region from the shocked wind material. This inner shock reaches its stagnation radius $\mathit{r}_\mathrm{stag}$ when the wind ram pressure $\mathrm{\rho}(r)v_\mathrm{w}^2$ becomes equal to the external ambient medium pressure $P_\mathrm{{amb}}$, so

\begin{equation}\label{stagrad}
\begin{split}
& \mathit{r}_\mathrm{{stag}} = \sqrt{\frac{\dot{M}\mathrm{_w}v\mathrm{_w}}{4\pi P\mathrm{_{amb}}}} \approx \\
&\approx 6.3\times10^{13}\left(\frac{\dot{M}\mathrm{_w}}{10^{-7} \mathrm{M_{\odot} /yr}}\right)^{1/2}\left(\frac{v\mathrm{_w}}{50\mathrm{\; km/s}}\right)^{1/2}\left(\frac{P\mathrm{_{amb}}}{6.2\times10^{-4} \mathrm{dyn/cm^2}}\right)^{-1/2} \mathrm{cm},
\end{split}
\end{equation}

where ``standard'' values for the wind parameters (see Section \ref{results}) and a reference value for the hot environment thermal pressure at pericenter are adopted. Due to the high pressure, the stagnation radius is very small and the shocked wind material forms a dense and rapidly cooling thin shell.
Another main difference with respect to the classic wind structure is that the winds considered here have subsonic expansion velocities. So, in the ideal case of a source at rest, in addition to the dense wind shell, a very weak sound wave propagates outward, with no significant reaction from the ambient medium on the wind. This case is not always true when the orbital velocity of the source is added to the expansion velocity, as we will discuss later.
It is also worth mentioning that the dense shocked wind shell is strongly subject to the Rayleigh-Taylor instability (RTI). In fact, in the frame of reference of the contact discontinuity between the wind material and the hot atmosphere, there is a static dense medium (the shocked wind shell) and a lighter one (the atmosphere) accelerated toward it. An order of magnitude calculation of the RTI timescale results in a much shorter value than G2's orbital period. 
In summary, the shape of stellar winds changes significantly when the winds are located in a high-pressure environment.
For further details, we refer the reader to \citet{Parker_63}, \citet{Koo_92}, \citet{vanMarle_06} and A. Ballone et al. (in preparation).

\begin{figure}
\begin{center}
\includegraphics[scale=0.7]{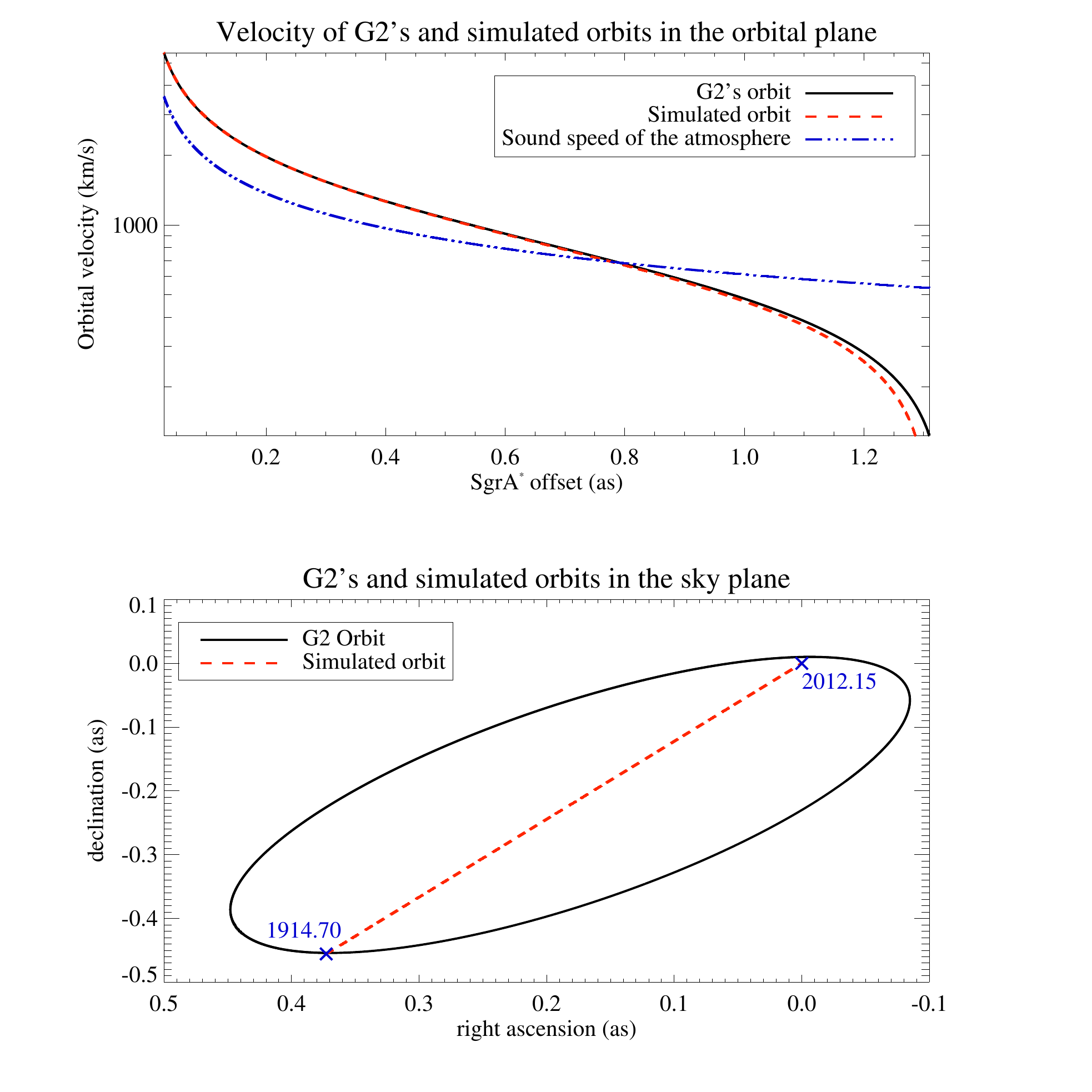}
\caption{Comparison of G2's best-fit orbit (black solid line) and the simulated $e=1$ orbit (red dashed line). The upper panel shows the evolution of the total velocity along the orbits. The blue dash-dotted line shows the sound speed of the atmosphere along the orbit. The lower panel shows the two orbits when projected on the sky plane. The blue crosses show the simulated orbit's apocenter and pericenter, with their relative orbital time.
}\label{orbits}
\end{center}
\end{figure}

Considering the effects of the motion of the source is also important. \citet{Burkert_12} have shown that near pericenter the ram pressure of the hot atmosphere becomes comparable to (and eventually even higher than) its thermal pressure, so one expects stripping of the wind material. The ram pressure will also give an additional "non-thermal" contribution to $P\mathrm{_{amb}}$ in Equation (\ref{stagrad}), with a different intensity depending on the angle $\theta = \mathbf{v}_\mathrm{w}/\mathbf{v}_*$ (with $\mathbf{v}_*$ being the wind source velocity along the orbit, see Figure \ref{orbits}). Furthermore, even if the expansion velocity of the wind is subsonic, at a distance of roughly $\mathrm{0''.8\approx 10^{17}\; cm}$ from SgrA* the orbital velocity of the source becomes slightly supersonic (Figure \ref{orbits}) and, at that point, an external weak bow shock forms in the ambient medium.

Finally, one should also consider the effect of the strong gravity G2 is subject to. As already discussed by \citet{Schartmann_12}, tidal forces become dominant near G2's closest approach to the SMBH. As shown in Section \ref{results}, the tidal force leads to the stretching of the wind shells, dramatically affecting the global structure of the outflow near G2 pericenter, as well as its distribution in the position-velocity (PV) space.

\section{Physical and numerical setup}\label{setup}

To simulate single winds moving in the Galactic Center atmosphere we use two-dimensional cylindrical coordinates $(Z,R)$ in a rectangular grid with uniform (and equal) resolution in both coordinates (see Table \ref{param} for further details). The choice of two-dimensional simulations is mainly dictated by the need of extremely high resolution. We will show in Section \ref{results} that the structure of the emitting material is highly filamentary, which forced us to use resolutions that are roughly a factor of 10 higher compared with previous works in the so-called diffuse cloud scenario \citep{Schartmann_12, Anninos_12}. This work represents the first paper attempting detailed hydrodynamical simulations for the ``compact source'' scenario and trying to span the wind's mass-loss rate and velocity parameter space. Given the quite high degree of complexity of the problem, several tests and studies have been performed, which led to a large computational effort and which could not be done with three-dimensional simulations. Different resolutions have been used for the different models and, because of the high computational cost, we decided not to have more than $2\times10^7$ grid cells. A reflective boundary is applied to the axis of symmetry, while outflow boundaries are set elsewhere. Given the extremely steep gradients of density and temperature of the atmosphere and of the velocity of the source along the orbit, we simulate almost the entire domain of the orbit from G2's apocenter, at $Z=-1.64\times10^{17}\mathrm{\;cm}$ \citep[from the orbital derivation of][]{Gillessen_13}, to very close to SgrA*, fixing the frame of reference on the SMBH. Our choice of cylindrical coordinates allows us to more correctly reproduce the spherically symmetric fluxes of the wind and, at the same time, to better simulate the quasi-axisymmetric interaction with the ambient medium without being forced to do computationally expensive three-dimensional simulations. On the other hand, this procedure led us to simplify our problem, assuming a zero angular momentum orbit. This restriction is not severe when the source is far enough from pericenter, since the observed orbit has a very high eccentricity $e\;\mathrm{\simeq0.97-0.98}$ \citep{Gillessen_13, Phifer_13}. In the top panel of Figure \ref{orbits} we plot a comparison of the total orbital velocity for the observed and simulated orbits, showing that the two curves depart from each other just near apocenter, at distances larger than $\approx 1'' \approx 1.25 \times 10^{17} \mathrm{cm}$ from the SMBH (the source in the simulated orbit starts at rest), where the ram pressure contribution is anyway insignificant. However, the observed and simulated orbits look very different (see Figure \ref{orbits}) when projected on the plane of the sky. The orbital time of the simulated orbit is roughly shifted by 1 yr compared with that of the observed orbit. For this reason, a strict comparison of times and ``projected'' quantities can be misleading. Therefore, we will often refer to distances to the SMBH rather than times, which are only roughly comparable to the observed times. 

In this work, we are only interested in modeling the head component, namely G2, so the wind parameters are chosen according to this choice.
The wind outflows are modeled with a mechanical approach, i.e., we fix a constant density and velocity in a very small circular input region in order to satisfy Equation (\ref{densform}) on its outer boundary. This condition holds as long as the sound speed of the injected material is negligible with respect to the wind velocity, which constrains the value of the temperature we can set in the input region. 
 In all our models, we set the temperature to a constant value $T\mathrm{_{in}}$ and an adiabatic index $\Gamma=1$ is assumed. This choice is based on the assumption that the temperature of the wind material is given by photoionization equilibrium due to the surrounding young stars \citep[$T\mathrm{_{in}=10^4 \;K}$;][]{Gillessen_12,Murray-Clay_12,Schartmann_12} and by the fact that the shocked wind material is so dense and cools so fast that it can be treated as being isothermal \citep{Scoville_13}. So, we set $T\mathrm{_{in}=10^4 \;K}$ in all our simulations except for the lowest velocity (LV) model described in Subsection \ref{velstudy} (for which such a temperature would give a sound speed comparable to the wind velocity). 

We let our source start from the orbit's apocenter position and we calculate the location of the source with a simple Runge-Kutta fourth order method. The outflow is also started at apocenter, following the original idea of \citet{Murray-Clay_12}. If the object has been scattered on the observed orbit by one or more close encounters, we can expect any pre-formed outflow to be disrupted and mostly stripped in these events.

The hot atmosphere is modeled following the density and temperature distribution used by \citet{Schartmann_12}, i.e.

\begin{equation}\label{denatm}
n\mathrm{_{at}}=930 \left(\frac{1.4\times10^4 \;R_\mathrm{S}}{d_\mathrm{{SMBH}}}\right)\;\mathrm{cm^{-3}},
\end{equation}
\begin{equation}\label{tematm}
T\mathrm{_{at}}=1.2\times 10^8\left(\frac{1.4\times10^4 \;R\mathrm{_S}}{d_\mathrm{{SMBH}}}\right)\mathrm{\; K},
\end{equation}

where $R_\mathrm{S}$ is the SMBH's Schwarzschild radius and $d_\mathrm{{SMBH}}$ is the distance from SgrA*.
These profiles correspond to the advection-dominated accretion flow (ADAF) analytical approximation of \citet{Yuan_03}, matching the current \textit{Chandra} X-ray observations \citep{Baganoff_03} and radio rotation measure data \citep{Bower_03}. As already discussed by \citet{Schartmann_12}, this atmosphere is convectively unstable, so we followed the same numerical recipe to artificially stabilize it, with the help of a passive tracer $tr$ advected with the wind material; i.e. every cell, where less than $0.01\,\%$  of the gas is made out of original cloud material, is reset to the initial condition of the atmosphere. As a result of this resetting, the formation of a bow-shock in the outer atmosphere is suppressed. However, any effect of this suppression on the wind material would not be severe. This bow shock in the atmosphere would be adiabatic and weak, with a Mach number of roughly $M\simeq1.5$ (see Figure \ref{orbits}). The Rankine-Hugoniot conditions give an increase in density of a factor $\approx1.7$ and a same decrease in the velocity. Thus, the ram pressure in the shocked ambient medium would be just a factor $\approx1.7$ lower than the simulated one. The stagnation radius would instead not change, since the total pressure is conserved across the shock. For a further and more detailed discussion on the chosen atmosphere, we refer the reader to \citet{Burkert_12} and \citet{Schartmann_12}.

Finally, we included the SMBH's gravitational field, modeled as a Newtonian point-source with mass $M_\mathrm{{BH}}  = 4.31 \times 10^6 \; \mathrm{M_{\odot}}$ \citep{Gillessen_09} at $Z,R=0$. The gravity of the central object is not considered, since the Roche radius for a $1 \; M_{\odot}$ star on G2's orbit is always at least a factor of three smaller than the stagnation radius given in Equation (\ref{stagrad}), hence we do not expect any significant change of the structure of the wind shell due to the stellar gravity.

The hydrodynamical equations are solved using a piecewise parabolic method in space and a Runge-Kutta third order method in time. Fluxes are computed with the two-shock Riemann Solver \citep{Mignone_07}.

\begin{deluxetable}{l l l l l l}
\tabletypesize{\scriptsize}
\tablewidth{-1pt}
\tablecaption{Parameters of the simulated models.\label{param}}
\tablehead{
\colhead{} & \colhead{$\dot{M}\mathrm{_w} \mathrm{(M_{\odot} \;yr^{-1})}$} & \colhead{$v\mathrm{_w} \mathrm{(km/s)}$} & \colhead{resolution} & \colhead{domain size (Z $\times$ R) ($10^{16}$ cm)} & \colhead{n. of grid cells}
}
\startdata
standard model & $\mathrm{8.8 \times 10^{-8}}$ & 50 & $7.5\times10^{12}$ cm & $[-17.10:-0.30]\times[0:0.45]$ & $1.344\times10^7$\\
 & & & 0.5 AU & & \\
LV & $\mathrm{8.8 \times 10^{-8}}$ & 10 & $7.5\times10^{12}$ cm &$[-17.10:-0.30]\times[0:0.45]$ & $1.344\times10^7$\\
 & & & 0.5 AU & & \\
HV & $\mathrm{8.8 \times 10^{-8}}$ & 250 & $2\times10^{13}$ cm &$[-18.00:-0.15]\times[0:3.00]$& $1.33875\times10^7$\\
 & & & 1.3 AU & & \\
LMDOT & $\mathrm{1.76 \times 10^{-8}}$ & 50 & $7.5\times10^{12}$ cm & $[-17.10:-0.30]\times[0:0.45]$& $1.344\times10^7$\\
 & & & 0.5 AU & & \\
HMDOT & $\mathrm{4.4 \times 10^{-7}}$ & 50 & $10^{13}$ cm & $[-18.00:-0.15]\times[0:1.00]$ & $1.785\times10^7$\\
 & & & 0.7 AU & & \\
LOWRES & $\mathrm{8.8 \times 10^{-8}}$ & 50 & $1.5\times10^{13}$ cm & $[-17.10:-0.30]\times[0:0.45]$&$3.36\times 10^6$\\
 & & & 1.0 AU & & \\
\enddata
\end{deluxetable}

\begin{figure}
\begin{center}
\includegraphics[scale=0.2]{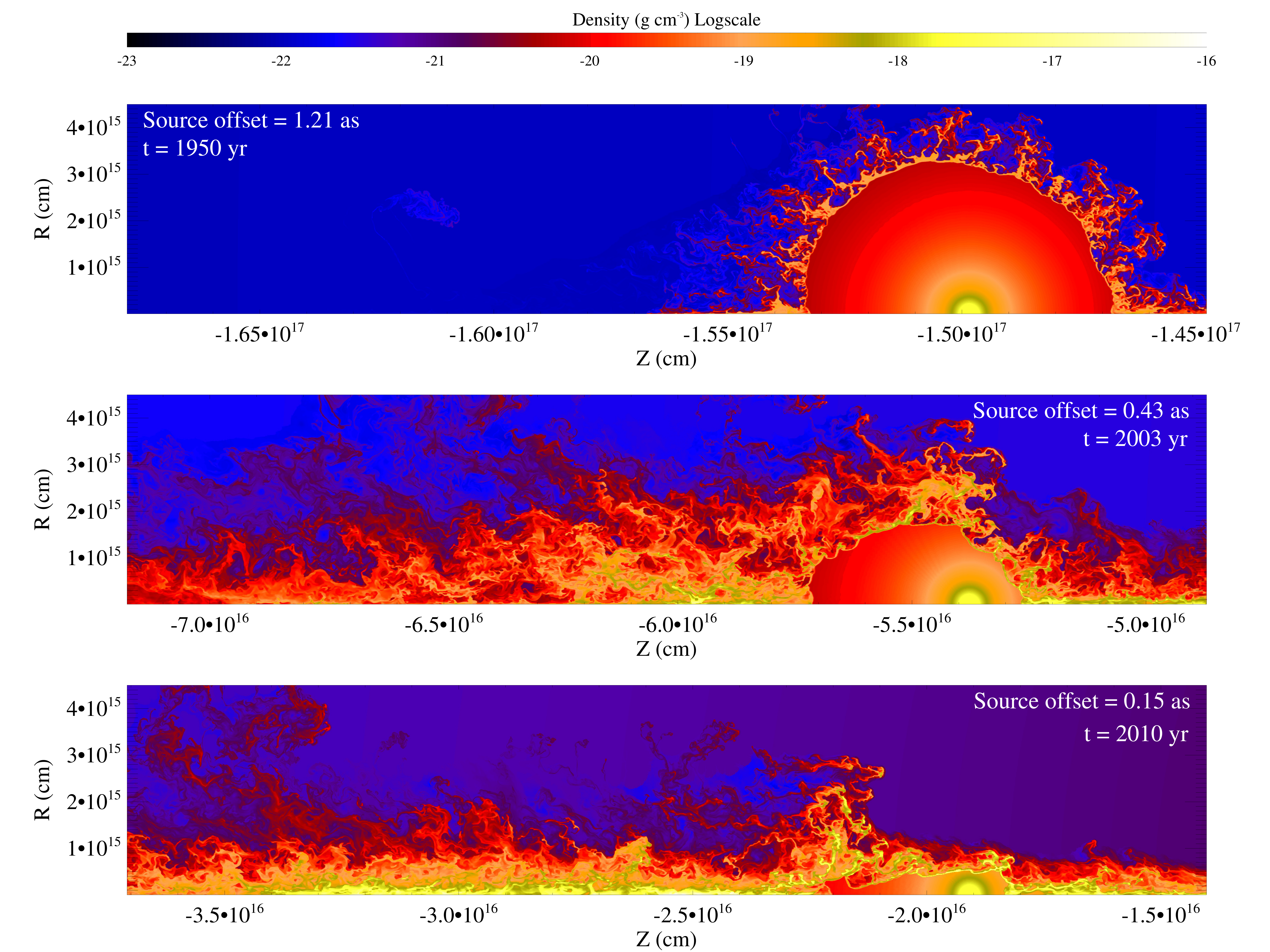}
\caption{Density maps for our standard model, for source distances of $1''.21, 0''.43$, and $\mathrm{0''.15}$ from SgrA* (from top to bottom).
}\label{bmden}
\end{center}
\end{figure}

\section{Results}\label{results}

In Subsection \ref{best} we present our standard model, discussing its evolution and main features and comparing it with the observations. In Subsections \ref{velstudy} and \ref{mdotstudy}, we discuss how the results change when the outflow velocity and mass-loss rate are varied, respectively.

\subsection{Standard model}\label{best}

\subsubsection{Evolution of the wind}

Our standard model has $\dot{M}\mathrm{_w=8.8\times 10^{-8} M_{\odot} \;yr^{-1}}$ and $v\mathrm{_w = 50 \;km\, s^{-1}}$. This value for the mass-loss rate is intended to reproduce the cloud mass estimated by \citet{Gillessen_12}. However, as we will show below, most of the luminosity of our winds comes from dense and filamentary material (see Subsection \ref{compobs}). So, in this case, the mass estimate of these authors, based on a constant density over an ellipsoidal volume, does not hold anymore. In other words, given the complex gas distribution, properties, and emissivities in our wind models, there is not a simple conversion between the total mass of the wind material and its luminosity. Regardless, the mass injected in the case of our standard model corresponds to the mass estimated by these authors.

In Figure \ref{bmden}, we show the density distribution of the wind (in just a fraction of our total two-dimensional computational domain) for three different positions along the orbit of the source. In these images, three different regimes are clearly visible. 

\begin{itemize}
\item In 1950, the wind is at a distance of $\mathrm{\approx 1''.21 \approx 1.5 \times 10^{17} \;cm}$ from the SMBH and in this part of the orbit the thermal pressure of the atmosphere is the main confinement affecting the outflow. Its structure is still almost spherical, with the free-wind region occupying most of the wind volume and the denser and very thin shocked wind shell that has already developed and turbulent Rayleigh-Taylor fingers departing from it. 
\item In 2003, at a distance of $\mathrm{\approx 0''.43\approx 5.4\times 10^{16}\;cm}$, the ambient ram pressure has generated a long tail of lower density stripped material that is mixing with the atmosphere. The density of the shocked wind material is now higher due to the increase of the thermal and ram pressure of the atmosphere while approaching SgrA*. This ram pressure and the tidal force of SgrA* has also broken the spherical symmetry of the free-wind region: the ram pressure has compressed the front part, leading to a smaller stagnation radius in the direction parallel to the motion (see Equation (\ref{stagrad})).  
\item The 2010 snapshot (corresponding to a distance of $\mathrm{\approx 0''.15 \approx 1.9\times 10^{16}\;cm}$) shows that the shocked wind material has reached even higher densities and that it has accumulated along the axis of symmetry due to the extremely high tidal stretching and compression in the direction perpendicular to the motion. A very tiny free-wind region is left at this time. 
\end{itemize}

\begin{figure}[!h]
\begin{center}
\includegraphics[scale=0.64]{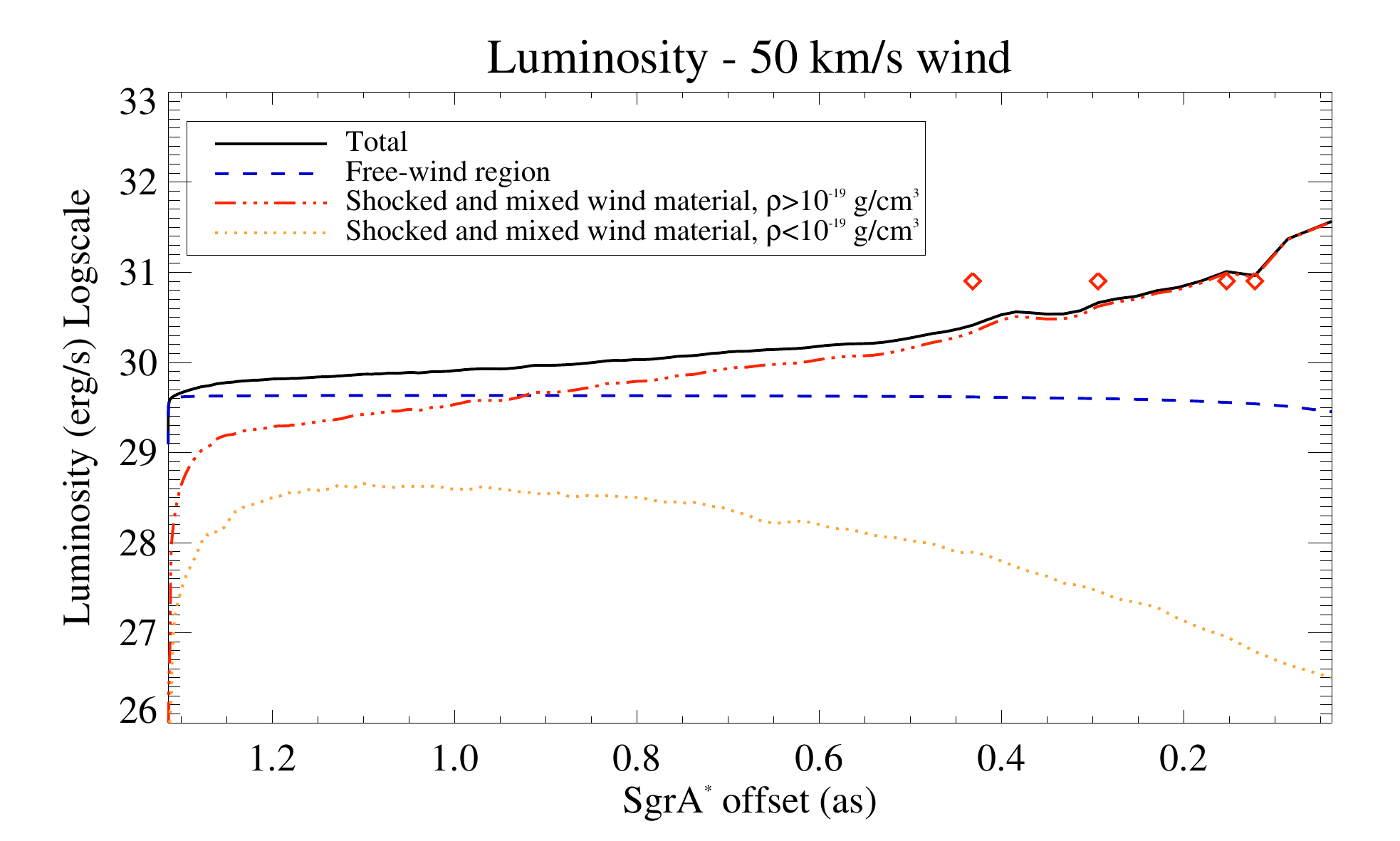}
\includegraphics[scale=0.21]{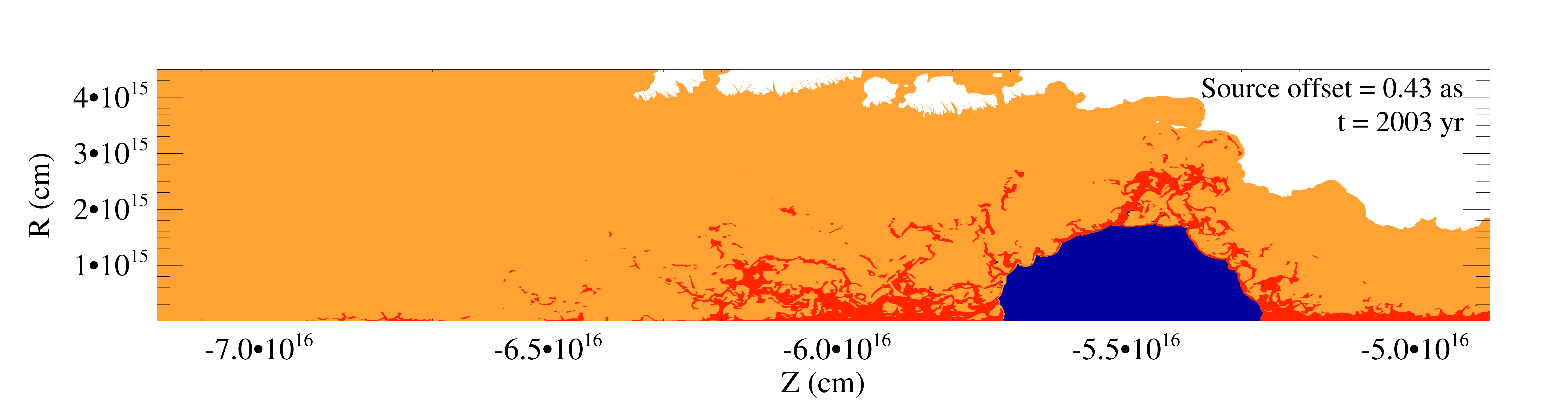}
\caption{Top panel: Br$\gamma$ luminosity evolution for our standard model with wind parameters $\dot{M}\mathrm{_w=8.8\times 10^{-8} \;M_{\odot}\;yr^{-1}}$ and $v\mathrm{_w=50\; km\;s^{-1}}$. The black solid line shows the total luminosity, the blue dashed line shows the luminosity of the free-wind region, the red dash-dotted line shows the luminosity of the shocked wind material with densities higher than $10^{-19}\;\mathrm{g\;cm^{-3}}$, and the orange dotted line shows the luminosity of the shocked wind material with densities lower than $10^{-19}\;\mathrm{g\;cm^{-3}}$. The red diamonds represent the observations.
Bottom panel: wind material distribution for a source distance of $\mathrm{0''.43}$ from SgrA*. The colors are the same as in the top panel.}\label{lumbm}
\end{center}
\end{figure}

\subsubsection{Comparison with observations}\label{compobs}

In Figure \ref{lumbm}, we plot the Br$\gamma$ luminosity evolution along the orbit (top panel, black solid line). To calculate the evolution, we used a functional form for the case B recombination Br$\gamma$ emissivity 

\begin{equation}\label{lumform}
j\mathrm{_{Br\gamma}}= 3.44\times10^{-27}  (T/10^4\mathrm{\;K})^{-1.09}n_\mathrm{p}n\mathrm{_e \;erg\;s^{-1}\;cm^{-3}},
\end{equation}

where $T$ is the wind material temperature and $n_\mathrm{p}$ and $n_\mathrm{e}$ are the proton and electron number densities, respectively, obtained by extrapolating the values given on page 73 in \citet{Osterbrock_06}. We then integrated this form over all the wind material, i.e., over the grid cells with wind tracer $tr>10^{-4}$ (the first cell in the $R$ direction is excluded from this calculation, as explained and justified in Section \ref{numiss}). We took into account the different temperatures of the wind material due to mixing with the atmosphere. A limitation of this approach is that in our simulations the thermodynamics of the mixing is not modeled by detailed physics and it is simply given by hydrodynamical advection. When mixing with the hotter and less dense atmosphere, the wind material's emissivity significantly decreases. Assuming the former functional form thus provides much more realistic results.

For the part of the orbit covered by the observations (indicated by the red diamonds in Figure \ref{lumbm}), the luminosity ranges from a minimum value of 
$\mathrm{\simeq 2.59 \times 10^{30}\; erg/s \simeq 0.3} \;L\mathrm{_{Br\gamma,G2}}$ to a maximum value of $\mathrm{\simeq 9.06 \times 10^{30}\; erg/s} \simeq 1.2\; L_\mathrm{Br\gamma,G2}$.  These values are comparable with the observations, even if the luminosity of our standard model increases toward pericenter, while the observed one has a constant value of $L_\mathrm{Br\gamma,G2} \mathrm{\simeq 8 \times 10^{30}\; erg/s} \simeq 2 \times 10^{-3}\; L_{\odot}$ for the whole period covered by the observations \citep{Gillessen_13}.

One of the most interesting results is that - given our assumptions - most of the luminosity of the object results from the shocked wind material (red dash-dotted line in the top panel of Figure \ref{lumbm}), which has a very low volume filling factor (red area in the bottom panel of Figure \ref{lumbm}). A significant contribution to the luminosity could actually come from the very inner part of the free-wind region, where the density scales with $r^{-2}$. In our simulations, this region corresponds to our input region, where the density is fixed to a constant value. However, given the uncertainties in the ionizing process, the amount of ionized emitting material in this region is still a matter of debate \citep[see discussion in][]{Scoville_13}. In the protoplanetary disk model of \citet{Murray-Clay_12}, these authors assume a different density distribution for the ionized material in the disk, leading to a peak of emission at the disk edge, between roughly 10 and 50 AU. That peak would be just a local and minor peak if the \textit{shocked wind} material was taken into account. For a general wind solution, \citet{Scoville_13} also estimate that the cross section of the base of the wind (i.e., our wind input region) is too small to be ionized by the estimated flux of Ly$\alpha$ photons from the surrounding stars. On the other hand, our detailed simulations allowing for hydrodynamical instabilities and tidal stretching result in shocked wind material with a significantly larger cross section compared to the analytical estimates of \citet{Scoville_13}.

\begin{figure}[!h]
\begin{center}
\includegraphics[scale=0.27]{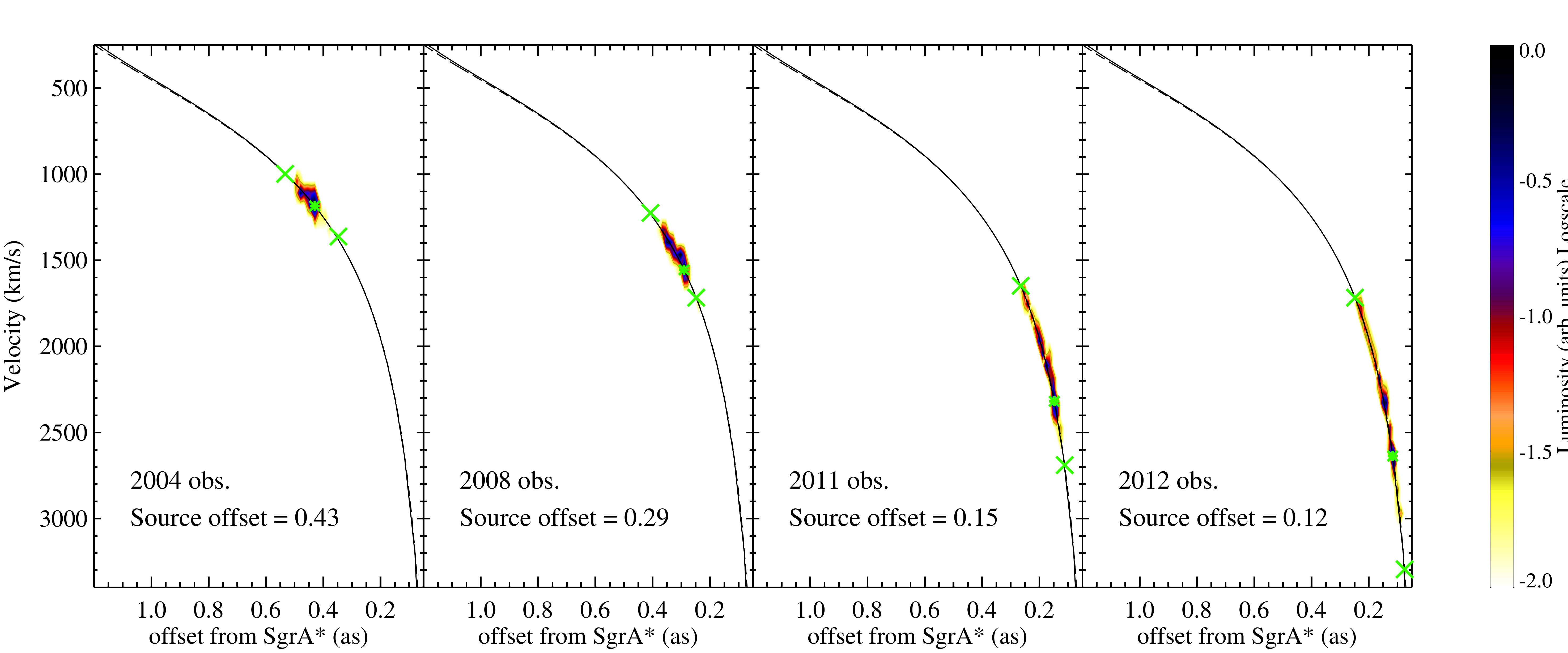}
\caption{Position-velocity diagrams for our standard model, for a source distance of $0''.43, 0''.29, 0''.15$, and $\mathrm{0''.12}$ from SgrA*. The green crosses show the G2 observed extremes and the green asterisk shows the position of the source in the diagram.
}\label{bmpv}
\end{center}
\end{figure}

PV diagrams for our standard model are shown in Figure \ref{bmpv}. As already discussed in Section \ref{setup}, the observed and the simulated orbits are very different when projected on the sky plane. For this reason, we plot velocity along the direction of the motion versus distance to the SMBH, instead of position and velocity projected on the sky. In order to compare these data with the observations, we deproject the G2 observed extremes along the orbit at different times and put these extremes on these plots (green crosses). As can be seen in Figure \ref{bmpv}, our standard model reproduces quite nicely the observed dynamical evolution of G2. Even if the match is not perfect, we will see in the next subsections that a slight variation of the parameters leads to significantly different sizes. 

Finally, as can be seen in both Figures \ref{bmden} and \ref{bmpv}, the source is never in the middle of the distribution, which shows that in the case of a compact source scenario, the stripping of the wind material and the tidal stretching of it can lead to a ``dynamical decoupling'' between the source and the extended emitting material, with the source being at late times much nearer the leading edge of the object. Hence, the orbit of the source itself can be slightly offset from the observed orbit, determined from the gas and dust emission.

\begin{figure}
\begin{center}
\includegraphics[scale=0.2]{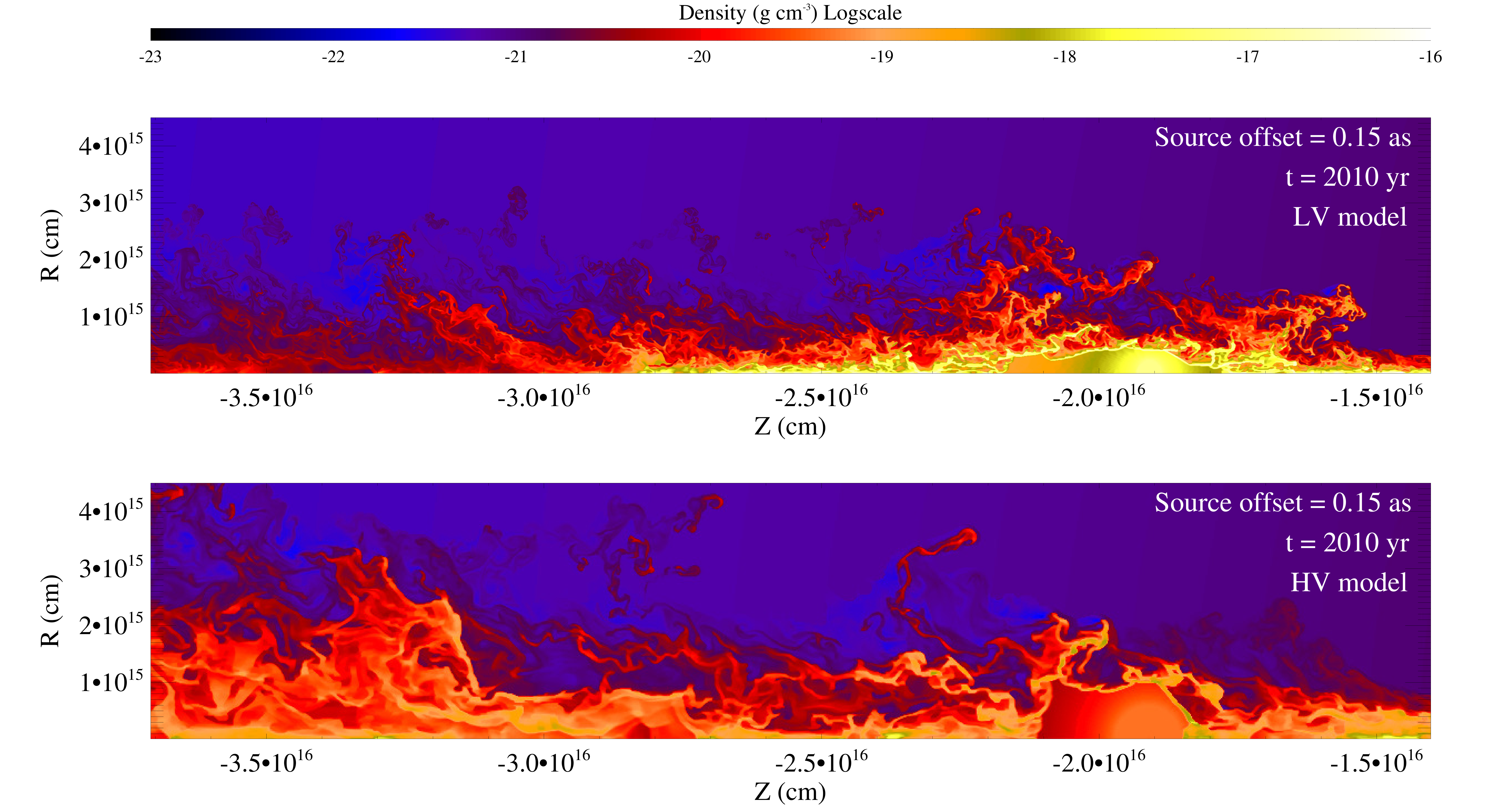}
\caption{Density maps for the $v\mathrm{_w=10 \;km\;s^{-1}}$ LV model (top panel) and $v\mathrm{_w=250 \;km\;s^{-1}}$ HV model (bottom panel), for a source distance of $\mathrm{0''.15}$ from SgrA*.
}\label{velpaden}
\end{center}
\end{figure}

\begin{figure}
\begin{center}
\includegraphics[scale=0.3]{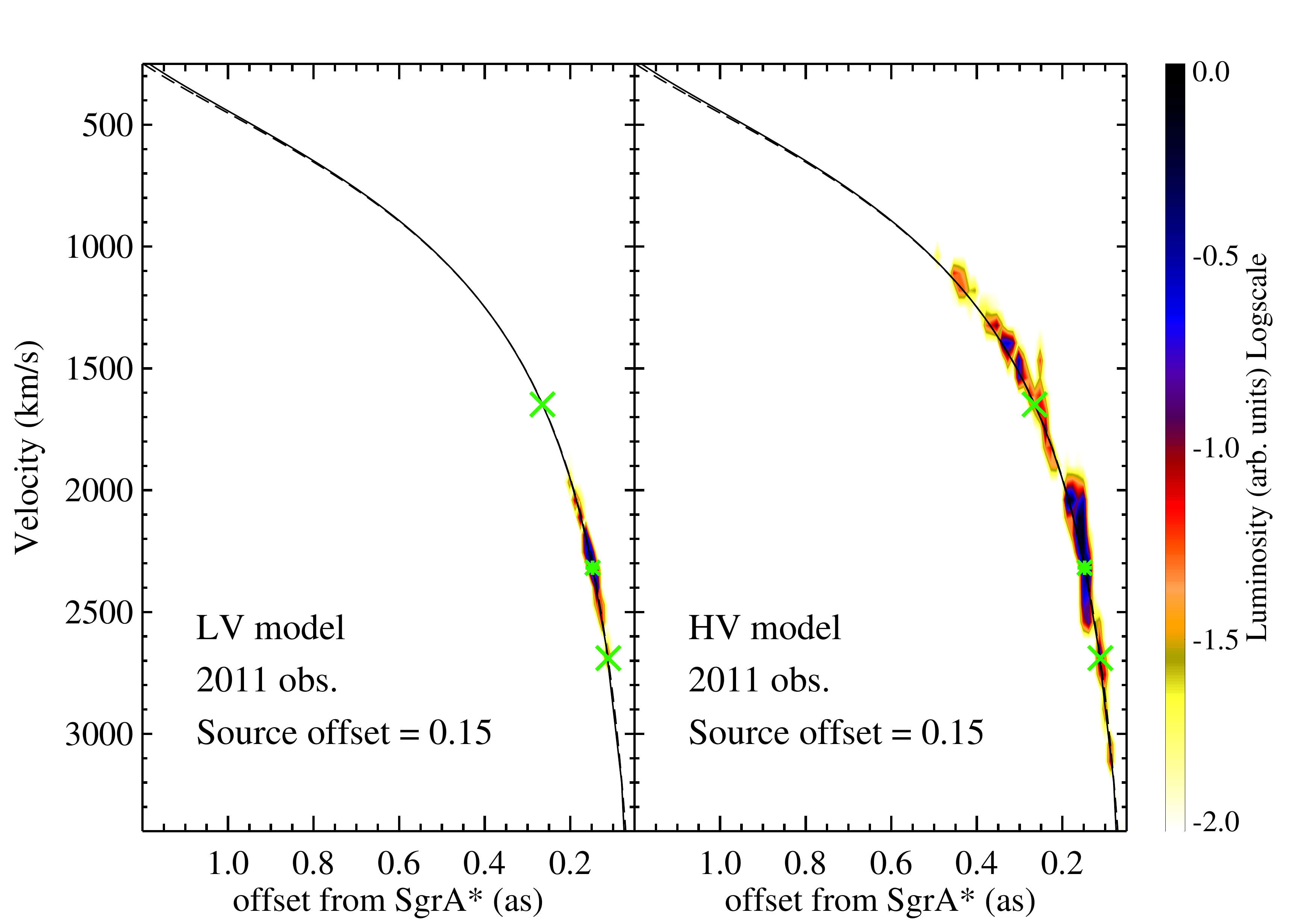}
\caption{Position-velocity diagrams for the $v\mathrm{_w=10 \;km\;s^{-1}}$ LV model (left panel) and $v\mathrm{_w=250 \;km\;s^{-1}}$ HV model (right panel), for a source distance of $\mathrm{0''.15}$ from SgrA*. The green crosses show the G2 observed extremes and the green asterisk shows the position of the source in the diagram.
}\label{velpapv}
\end{center}
\end{figure}

\subsection{An outflow velocity study}\label{velstudy}

We studied the effect of a variation of the wind velocity on the structure and observed properties of the wind. Keeping the mass-loss rate fixed to the value of our standard model, we reduced and increased the wind velocity by a factor of five, obtaining the LV model and the HV model, respectively (see Table \ref{param}).

\begin{figure}[!h]
\begin{center}
\includegraphics[scale=0.67]{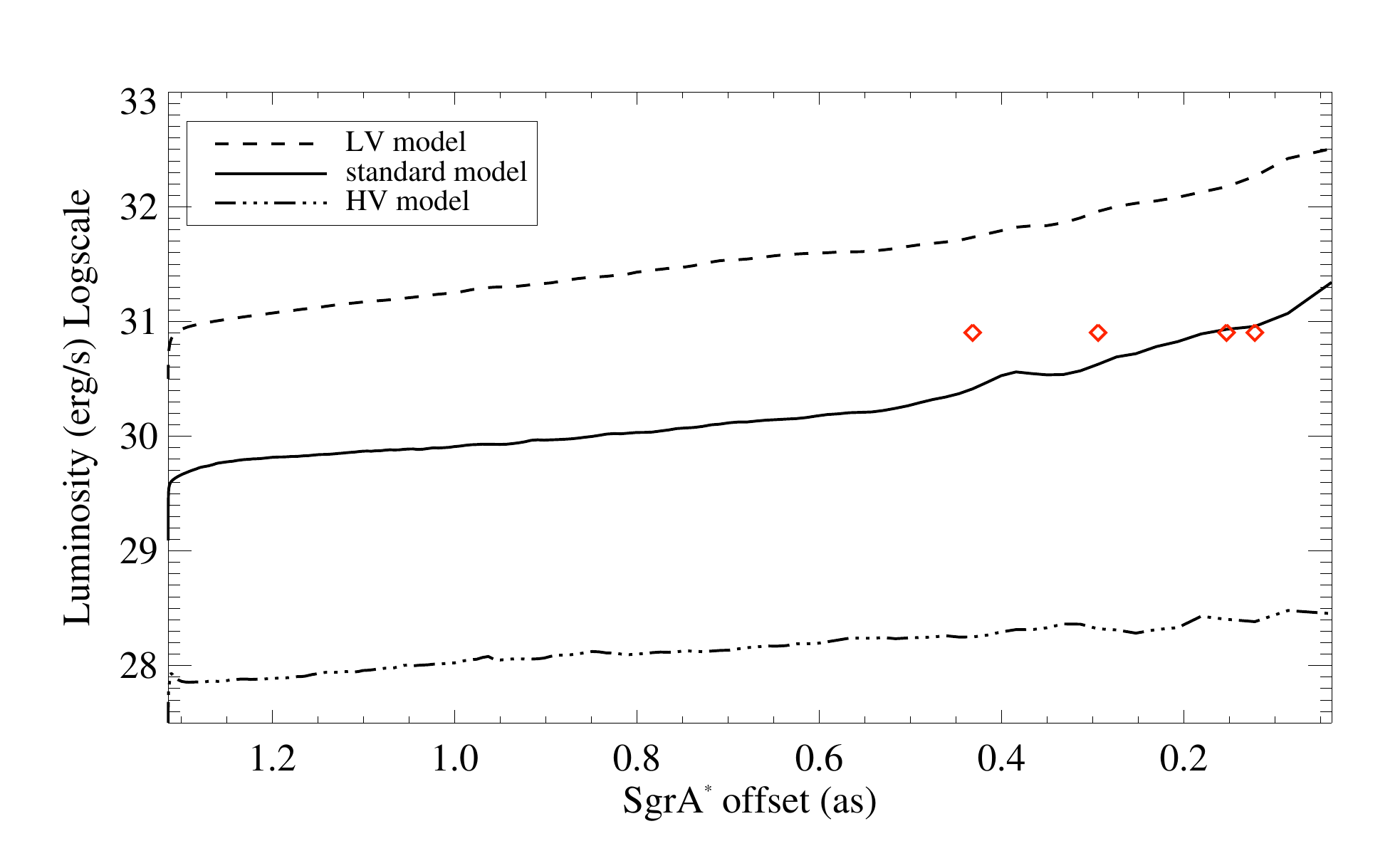}
\caption{Br$\gamma$ luminosity evolution for our wind velocity study. The standard model luminosity evolution is also included for a comparison. The red diamonds represent the observations.
}\label{vellum}
\end{center}
\end{figure}

In Figures \ref{velpaden} and \ref{velpapv}, we plot the density maps and the PV diagrams for a distance from SgrA* $\mathrm{\approx 0''.15}$, respectively. The density of the free-wind region and, hence, of the shocked wind material, is extremely different when changing the expansion velocity of the wind (see Equation (\ref{densform}) and the discussion in Subsection \ref{mdotstudy}). As a consequence, the amount of stripped and mixed material is much larger when the wind velocity increases. Both of these phenomena lead to a change in the Br$\gamma$ luminosity, as shown in Figure \ref{vellum}. The slower $\mathrm{10 \;km\;s^{-1}}$ LV model (dashed line) produces luminosities that are roughly one order of magnitude higher than those of our standard model (solid line). The faster $\mathrm{250 \; km\;s^{-1}}$ HV model (dash-dotted line) has instead roughly two orders of magnitude lower luminosities.
Significant differences are also visible in the PV diagrams: the LV model is $\approx \mathrm{0''.05}$ smaller than G2 (whose size is $0''.15$, at the considered position along the orbit) while the HV model has a much larger extension of $\approx \mathrm{0''.45}$, clearly exceeding G2's size. In terms of velocity dispersion, the HV model also shows a larger spread in velocity at given position, resulting from the higher wind velocity and from the higher turbulence of the stripped shocked wind material.

\subsection{An outflow mass-loss rate study}\label{mdotstudy}

We have also studied the effect of a variation of the wind mass-loss rate. In this case, we fixed the velocity to that of the standard model and reduced and increased the wind mass-loss rate by a factor of five; we obtained the LMDOT model and the HMDOT model, respectively (see Table \ref{param}).

\begin{figure}
\begin{center}
\includegraphics[scale=0.2]{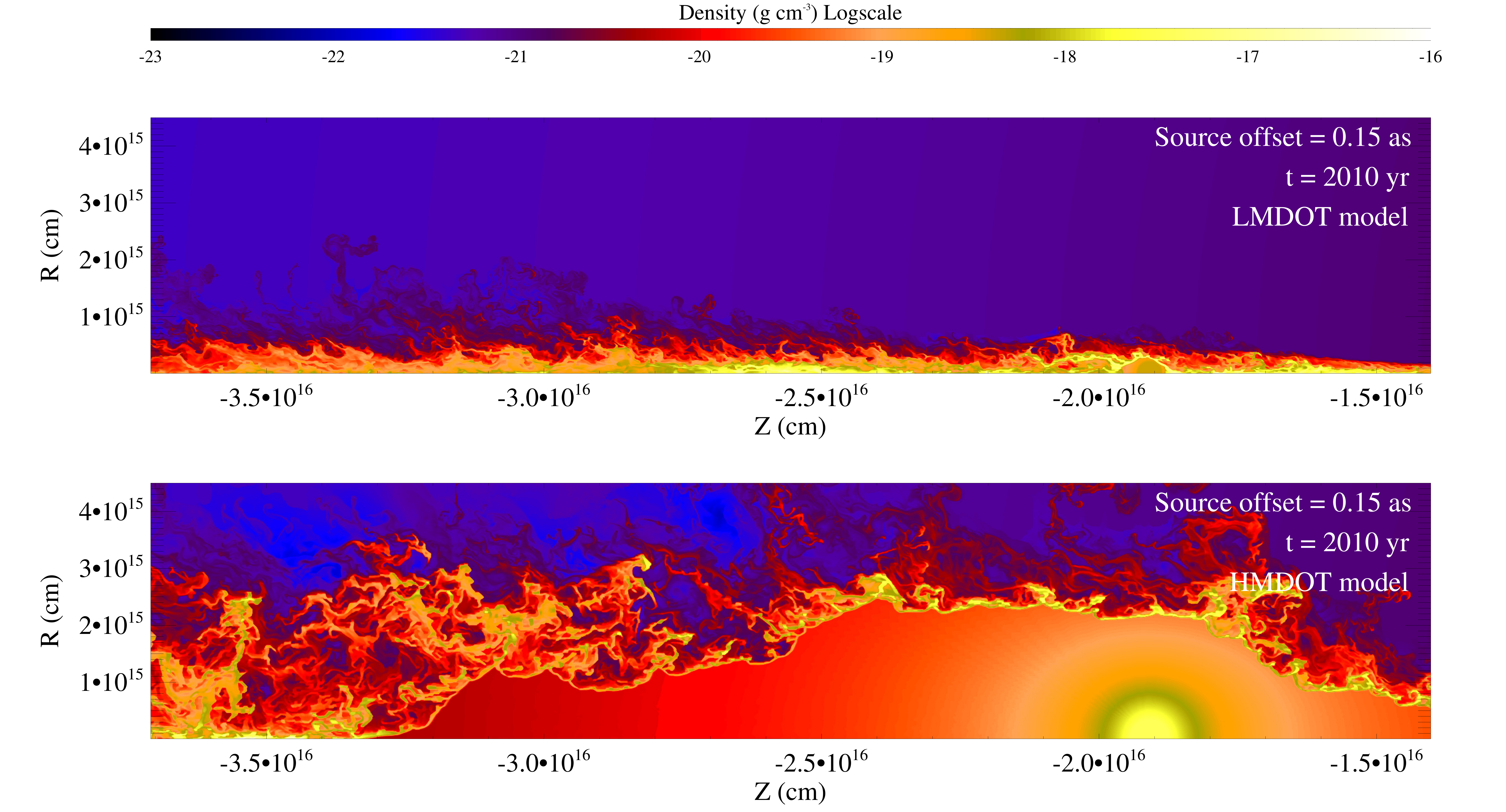}
\caption{Density maps for the $\dot{M}\mathrm{_w=1.76\times 10^{-8} \;M_{\odot} \;yr^{-1}}$ LMDOT model (top panel) and the $\dot{M}\mathrm{_w=4.4\times 10^{-7} \;M_{\odot} \;yr^{-1}}$ HMDOT model (bottom panel), for a source distance of $\mathrm{0.''15}$ from SgrA*.
}\label{mdotpaden}
\end{center}
\end{figure}

\begin{figure}
\begin{center}
\includegraphics[scale=0.3]{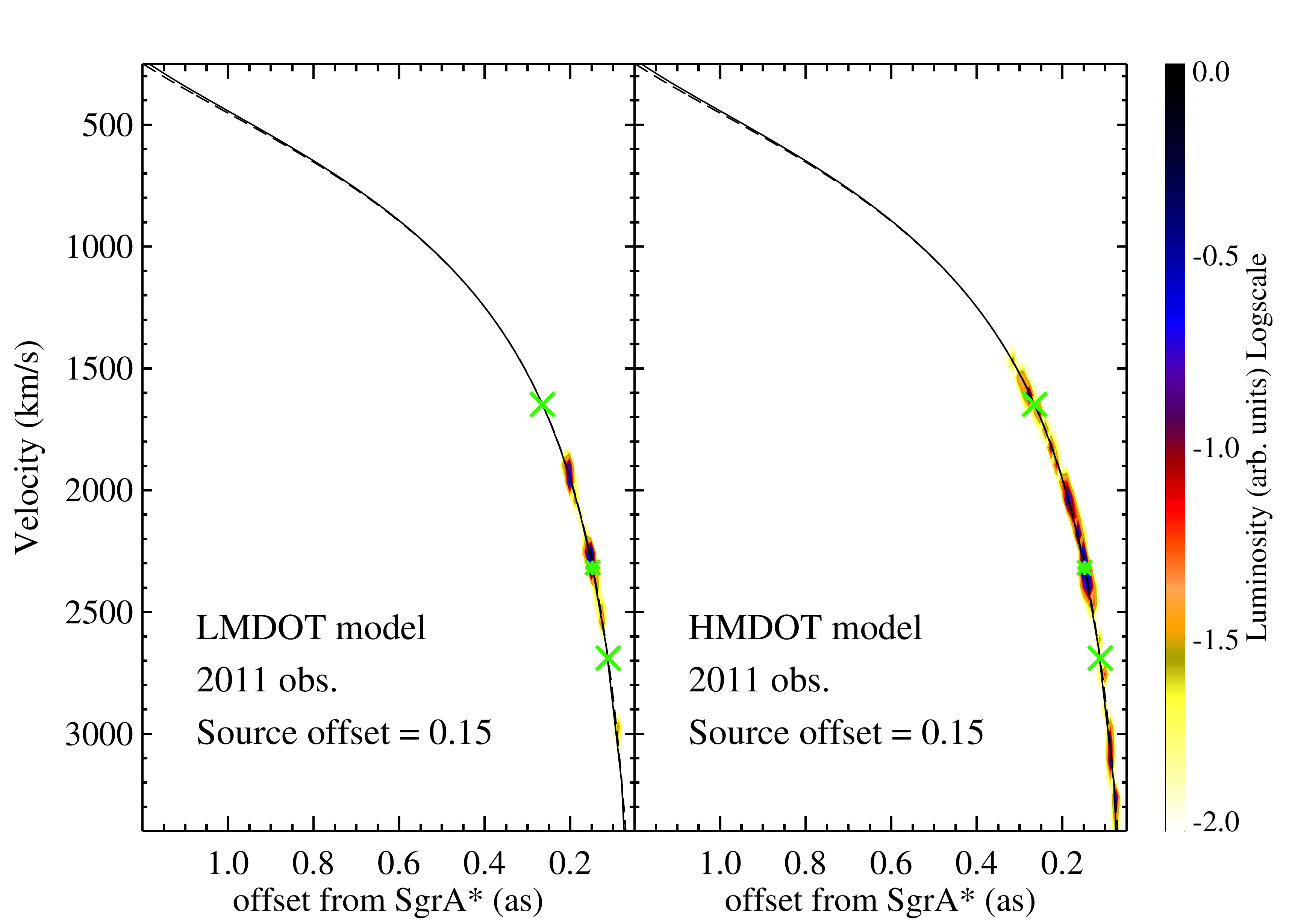}
\caption{PV diagrams for the $\dot{M}\mathrm{_w=1.76\times 10^{-8} M_{\odot} \;yr^{-1}}$ LMDOT model (left panel) and $\dot{M}\mathrm{_w=4.4\times 10^{-7} M_{\odot} \;yr^{-1}}$ HMDOT model (right panel), for a source distance of $\mathrm{0''.15}$ from SgrA*. The green crosses show the G2 observed extremes and the green asterisk shows the position of the source in the diagram.
}\label{mdotpapv}
\end{center}
\end{figure}

The density maps and the PV diagrams for the two different $\dot{M}\mathrm{_w}$ models, at a distance from SgrA* $\mathrm{\approx 0''.15}$, are shown in Figures \ref{mdotpaden} and \ref{mdotpapv}, respectively. 

\begin{figure}[!h]
\begin{center}
\includegraphics[scale=0.2]{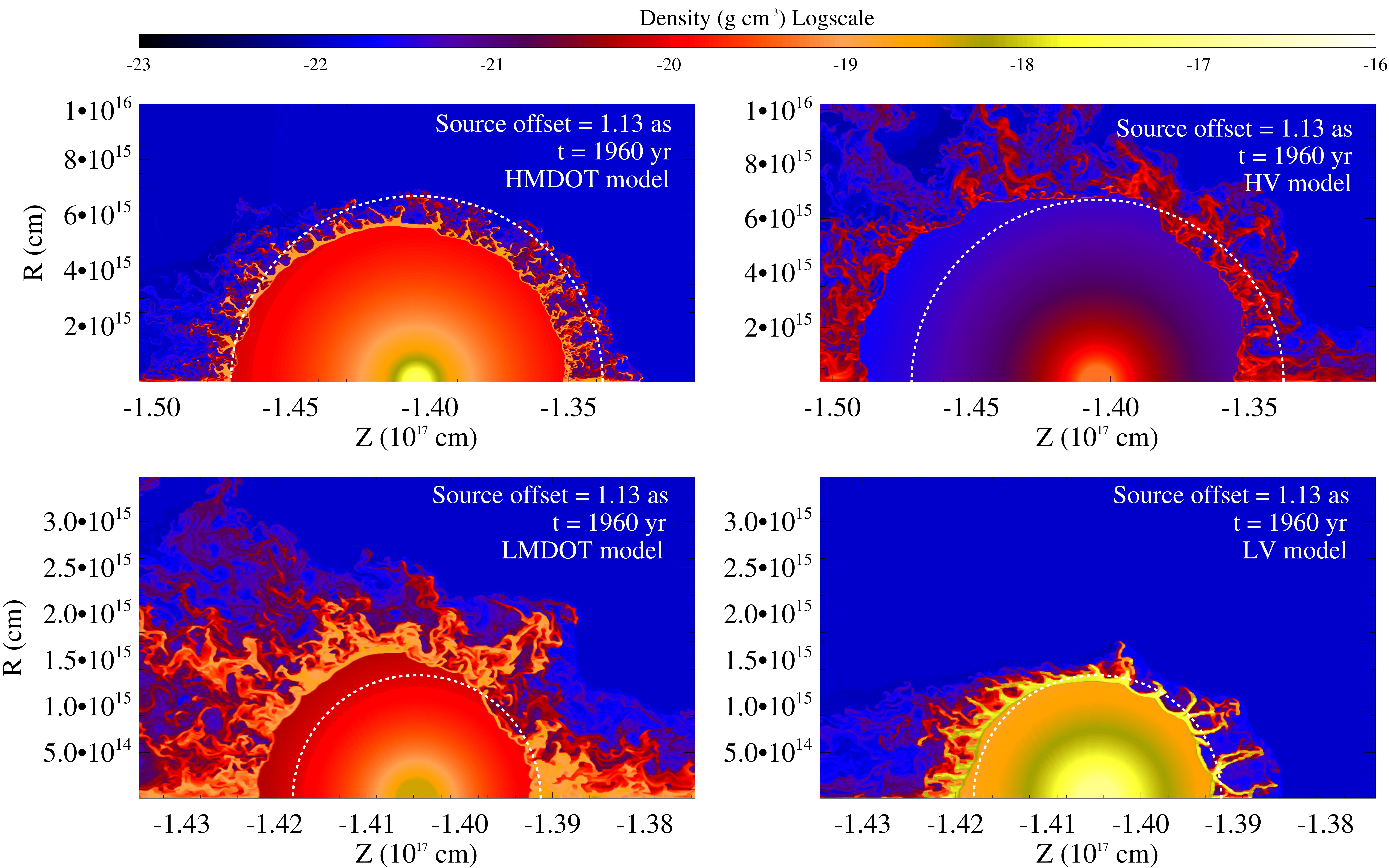}
\caption{Density maps for the HMDOT and HV models (top panels) and the LMDOT and LV models (bottom panels), for a source distance of $\mathrm{1''.13}$ from SgrA*. The white dotted line denotes the expected theoretical size of the stagnation radius when taking into account only the thermal pressure of the atmosphere at that distance from SgrA*.
}\label{staglow}
\end{center}
\end{figure}
A first inspection of the density and luminosity distribution necessitates a discussion of the structure of the winds at late times. In fact, given Equation (\ref{stagrad}), we expect the size of the free-wind region to be equal in the case of the LV and LMDOT models and similarly in the case of the HV and HMDOT models, given the same $\dot{M}\mathrm{_w}v\mathrm{_w}$. Figure \ref{staglow} shows that this fact is indeed true for earlier times; this figure provides a first qualitative explanation for the different sizes visible in Figure \ref{mdotpapv}, but the stagnation radius equation does not strictly apply at late times. An explanation for this behaviour is mainly given by considering the different impact of the ram pressure of the atmosphere, which depends on the densities of the different models. For different models, an equal $\dot{M}\mathrm{_w}v\mathrm{_w}$ means an equal wind ram pressure $\rho\mathrm{_w}(r) v\mathrm{_w}^2$. So, a factor of five lower velocity at constant $\dot{M}\mathrm{_w}$ implicitly implies a factor of five higher $\rho\mathrm{_w}(r)$ and vice versa. On the other hand, at constant $v\mathrm{_w}$, a factor five lower $\dot{M}\mathrm{_w}$ implies a factor five lower $\rho\mathrm{_w}(r)$ and vice versa. The difference in density between the LV and LMDOT models (and the HV and HMDOT models) is then a factor 25. Obviously, this difference also affects the density of shocked wind material. We can therefore distinguish two regimes and see that in the case of the LMDOT and HV models, given the global lower densities, the ram pressure stripping of the atmosphere acts much more efficiently, reducing the size of the free-wind region at late times and accumulating backflowing stripped material behind the source. For the LV and HMDOT models, instead, the stripping is less efficient and the size of the free-wind region at late times is mainly given by the tidal stretching.
In addition, Figure \ref{staglow} also shows that, for the same $\dot{M}\mathrm{_w}v\mathrm{_w}$, the higher velocity models (namely LMDOT and HV) have more elongated and turbulent RTI fingers, increasing the wind cross section. This phenomenon occurs because, at fixed wind ram pressure, faster winds have lower momentum, so they experience higher deceleration due to the external pressure, i.e., the higher the velocity of the wind, the more quickly the stagnation radius reached. The typical timescale for the growth of RTI is inversely proportional to the square root of the acceleration, meaning that winds with higher velocities have more unstable shells.

\begin{figure}[!h]
\begin{center}
\includegraphics[scale=0.67]{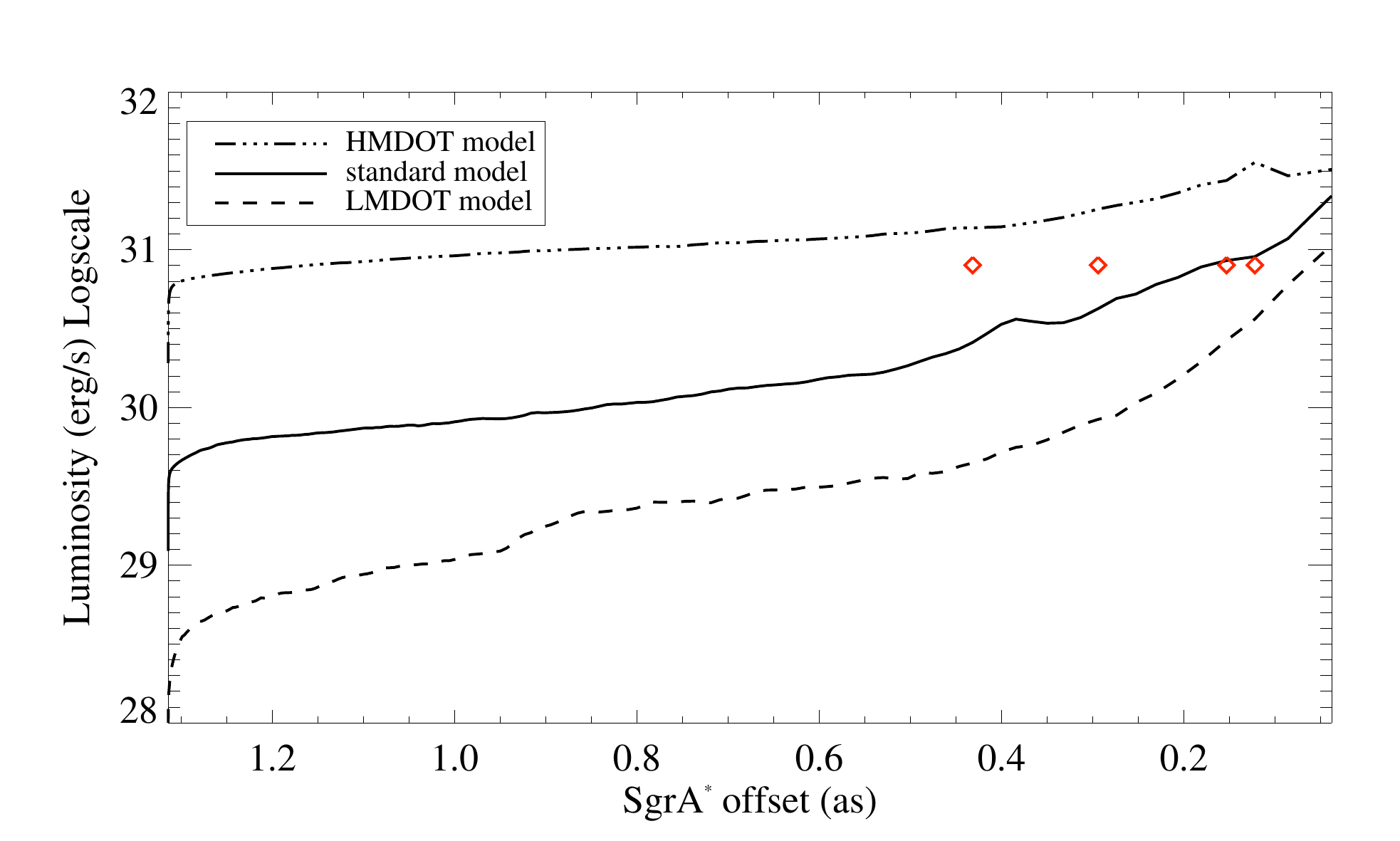}
\caption{Br$\gamma$ luminosity evolution for the $\dot{M}\mathrm{_w=1.76\times 10^{-8} \;M_{\odot} \;yr^{-1}}$ LMDOT model and for the $\dot{M}\mathrm{_w=4.4\times 10^{-7} \;M_{\odot} \;yr^{-1}}$ HMDOT model. The standard model luminosity evolution is also included for a comparison. The red diamonds represent the observations.
}\label{mdotlum}
\end{center}
\end{figure}

Tidal stretching also plays a role in explaining the evolution of the luminosity. As seen in Figure \ref{mdotlum}, the evolution depends on $\dot{M}\mathrm{_w}$, with the slope of $L\mathrm{_{Br\gamma}(t)}$ increasing with decreasing mass-loss rate: for the part of the orbit corresponding to the observations, the luminosity increases by roughly a factor 2.6, 3.5 and 8.2, respectively, in the case of the HMDOT, the standard, and the LMDOT models. In the late phases, in fact, the tidal forces always compress and squeeze the wind towards the axis of the motion that corresponds to the axis of symmetry of our cylindrical coordinates. In other words, in the proximity of the SMBH, the stagnation radius in the direction perpendicular to the orbital motion is defined by the balance between the wind ram pressure and the pressure of the tidal forces. The decrease of the stagnation radius in this direction due to tidal compression will be evidently lower in the case of higher wind ram pressures. As a consequence, the increase of the density and luminosity of the shocked wind shell will also be different. This effect can be also partially recovered in Figure \ref{vellum} for our velocity study.

\section{Discussion}\label{discus}

\subsection{Numerical issues}\label{numiss}

It is well known that cylindrical coordinates lead to numerical artefacts near the axis of symmetry \citep[see, e.g.,][]{Vieser_07,Kwak_11}. In particular, in our case, all our models suffer from the formation of too elongated Rayleigh-Taylor fingers (emanating from the shocked wind shell) along this numerically critical part of the computational domain \citep[for a similar behaviour, see, for example, the hydrodynamical simulations of][]{Cox_12}. This problem could affect our results mainly in the direction of the leading part of the wind, where the ram pressure of these fingers seems to be artificially too high, thus reducing effective compression and stripping there. Quantifying the exact impact of these artificial features on the global evolution is rather difficult, but we believe that, given their narrowness, they should not have a strong effect on the stripping of the wind at larger $R$ positions. At the same time, considering their relatively small volume and mass (always less than 5\% of the total volume and mass), these numerical features do not contribute a strong weight to the total luminosity. Only a very dim artificial component, corresponding to the leading part of the object, appears in the PV diagrams, i.e., in the PV diagrams of the HV and LMDOT models (see Fig. \ref{velpapv} and \ref{mdotpapv}).
The cylindrical coordinates in combination with the reflective boundary at the axis of symmetry can lead to an artificial overcompression of the material in the first very few (one or two) cells at low $R$ values. For this reason, we excluded the first cell in our calculation of the luminosity. This choice does not change significantly our results, but allows us to remove some artificial and transient peaks in the luminosity evolution.

\begin{figure}[!t]
\begin{center}
\includegraphics[scale=0.2]{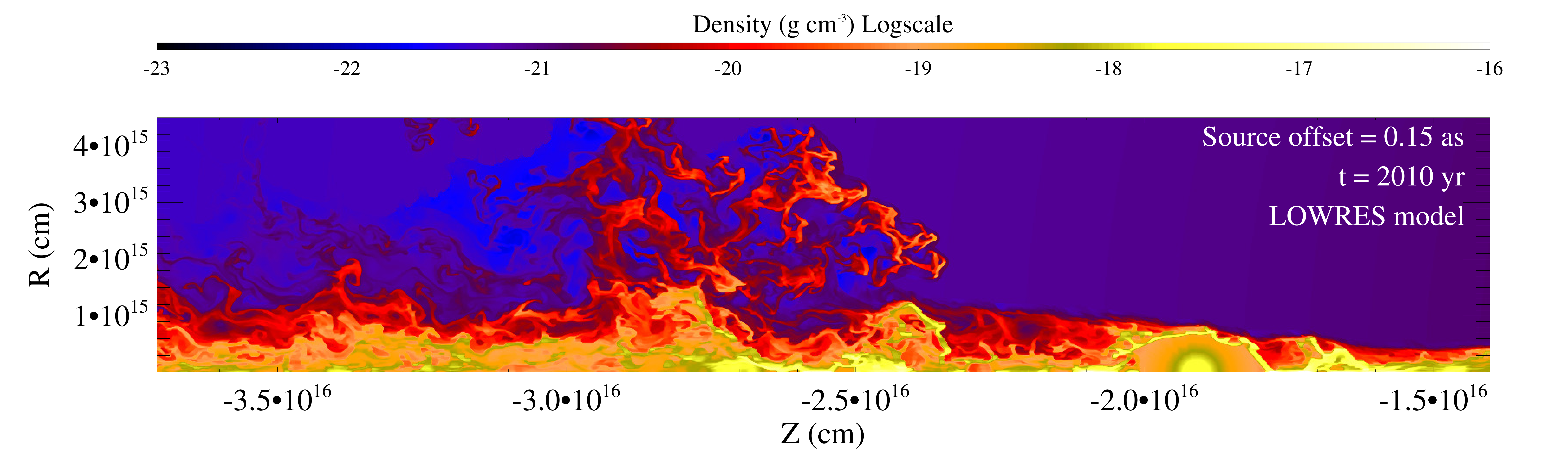}
\caption{Density map for our low resolution model, for a source distance of $\mathrm{0''.15}$ from SgrA*.
}\label{resden}
\end{center}
\end{figure}

\begin{figure}[!h]
\begin{center}
\includegraphics[scale=0.62]{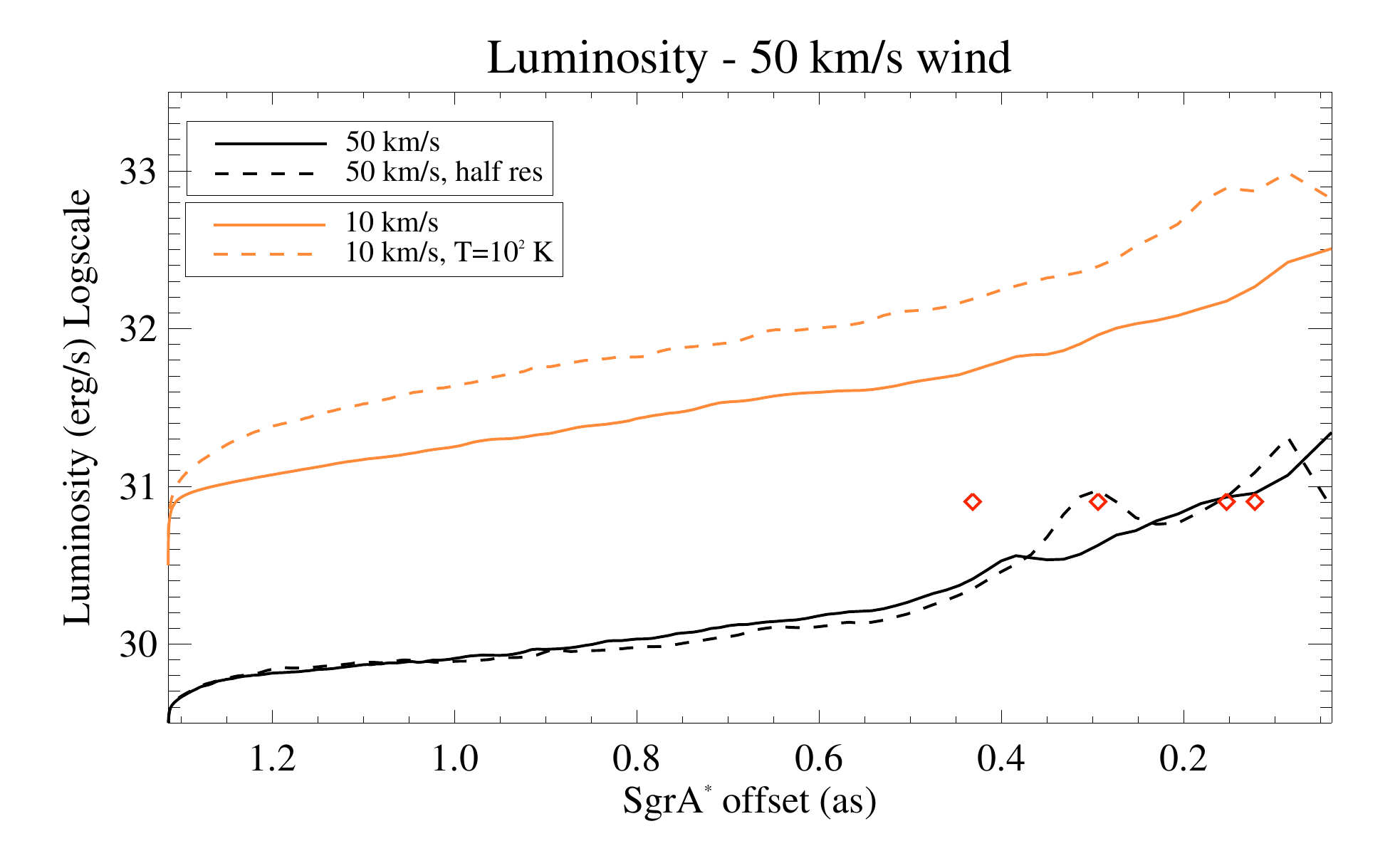}
\caption{Br$\gamma$ luminosity evolution for the tests discussed in Section \ref{numiss}. The solid black line shows our standard model, while the dashed black line shows its lower resolution counterpart. The solid orange line shows our LV model and the dashed orange line shows the same model with a temperature $T\mathrm{_{in}= 10^2 \; K}$. The red diamonds represent the observations.
}\label{lumresandt}
\end{center}
\end{figure}

The next two problems are related to the shell of shocked wind material.
First of all, computational cells that are too large are expected to lead to a poor resolution of the very thin shocked wind material shell, particularly in the later phases, when the external ambient thermal and ram pressure increase. For this reason, we used a very high resolution for our simulations (see Table \ref{param}). A further resolution increase would have led to an extremely high computational cost for any model, but we nonetheless checked how the resolution can affect the results. We thus doubled the grid cell size in all directions and ran again our standard model at this lower resolution (LOWRES model, see Table \ref{param}). To first order, the matter distribution is very similar to that of the standard model, as can be seen in Figure \ref{resden}. Also, the luminosity evolution is comparable (see Figure \ref{lumresandt}, black dashed line).

The second problem is related to the temperature of the shocked wind shell. In the LV model, the temperature of the wind is not set to $\mathrm{10^4\;K}$ (as in all the other models), but instead is set to $\mathrm{10^3\;K}$. The reason for this choice is related to our mechanical modeling of the winds. In the case of the LV model, the velocity we set in the input region roughly corresponds to the sound speed of a $\mathrm{10^4\;K}$ gas. Setting such a temperature in the input region would mean that the injected thermal and ram pressure becomes comparable, which would lead to an over-injection of mass and velocity (see Section \ref{setup}). However, a lower temperature in the wind material also leads to a lower thermal pressure in the shell of shocked wind material. This fact is a problem, because a lack of pressure support also produces slightly higher densities in this region and thus leads to higher luminosities (see Equation (\ref{lumform})). To get an idea of how this phenomenon can affect the resulting luminosities, we ran the LV model setting an even lower temperature of $\mathrm{10^2\;K}$. The luminosity evolution is shown in Figure \ref{lumresandt}: in this case, the luminosity is roughly a factor of three higher along most of the orbit. We therefore infer that the luminosity evolution of the LV model must be examined with care, since we cannot exclude systematically lower values in the case of a shell of temperature equal to $\mathrm{10^4\;K}$.

\subsection{Advantages, disadvantages and outlook for the compact source scenario} 

The compact source scenario is currently the main rival to the diffuse cloud scenario suggested and studied in great detail by \citet{Gillessen_12,Gillessen_13}, \citet{Burkert_12} and \citet{Schartmann_12}. The main problem of the diffuse cloud scenario remains the origin of G2. \citet{Burkert_12} have indeed shown that, given its observed evolution, a compact cloud must have formed around 1995 along the G2 orbit close to pressure equilibrium. However, no known star has been found close to G2's birthplace at that time and formation through a cooling instability of the hot atmosphere appears to be ruled out \citep{Burkert_12}.
The compact source scenario provides instead a plausible explanation for G2's origin, resulting from the scattering of a young low-mass star onto the current, highly eccentric orbit of G2, as discussed by \citet{Murray-Clay_12}, \citet{Miralda-Escude_12}, and \citet{Scoville_13}. The probability of such an event is very low, so in this case we should expect G2 to be a rare and peculiar object. On the other hand, if G2 is the outflow from a central source, we could expect this source to survive the close encounter with the SMBH and a new cloud could form after the disruption at the pericenter passage, leading to a ``periodic'' formation of G2. 

A possible candidate for G2's source, given our best parameters, could be a young T Tauri star, as suggested by \citet{Scoville_13}. These authors assumed $\dot{M}\mathrm{_w=4\times 10^{-8} M_{\odot} \;yr^{-1}}$ and $v\mathrm{_w = 100 \;km\;s^{-1}}$. Our standard (and best) model has roughly a factor two higher mass-loss rate and lower velocity. These values are a bit extreme, but still in the ranges of the observations \citep[$\dot{M}\mathrm{_w=[10^{-7},10^{-12}] \;M_{\odot} \;yr^{-1}}$ and $v\mathrm{_w = [50,300] \;km\;s^{-1}}$;][]{White_04}. T Tauri stars are young objects, with ages between $10^5$ and $10^7 \mathrm{yr}$ \citep[see, e.g.,][]{D'Antona_94}. This age is comparable with the age of the young stellar disk \citep{Paumard_06,Bartko_09} where this star was born and subsequently scattered, roughly $\mathrm{6\pm2 \;Myr}$ ago. A major caveat of this scenario is the geometry of the T Tauri outflows: there are several clues indicating that these outflows are bipolar winds or jets \citep[e.g.,][]{Hartigan_95,White_04}. In this work we decided to avoid adding further parameters to our study. It would be interesting, in the future, to see how a different geometry of the outflows affects the results.

Another advantage of the compact source scenario is that it could explain the tail (G2t) that is clearly visible in the 2011 and 2012 observations \citep{Gillessen_13}. As shown by \citet{Burkert_12} and \citet{Schartmann_12}, a large spherical shell at apocenter could explain both the observed head and tail. In this work, we focused on G2; however, a large shell resulting from a higher velocity wind in the compact source scenario could in principle behave in a similar way and give the tail the larger size and clumpy structure observed by \citet{Gillessen_13}. 

Three-dimensional simulations with a Cartesian grid would allow us to solve some of the numerical artifacts we discussed, remove the reduction to an $e=1$ orbit, and test the evolution of G2 on a complete orbit. These simulations, however, are very expensive in terms of computational time, so the two-dimensional simulations presented in this paper constitute a fundamental step to scan the available parameter space.

\section{Summary}\label{sumconc}

The aim of this work was to model, for the first time, the hydrodynamical evolution of a wind as it moves along G2's orbit, investigating its structure and observational properties as a result of the interaction with a hot atmosphere and the extreme gravitational field of the SMBH in the Galactic Center. For our study, we have been inspired by the compact source scenario suggested and studied by other authors \citep{Gillessen_12,Burkert_12,Murray-Clay_12,Miralda-Escude_12,Meyer_12,Scoville_13}.

\begin{figure}[!h]
\begin{center}
\includegraphics[scale=0.32]{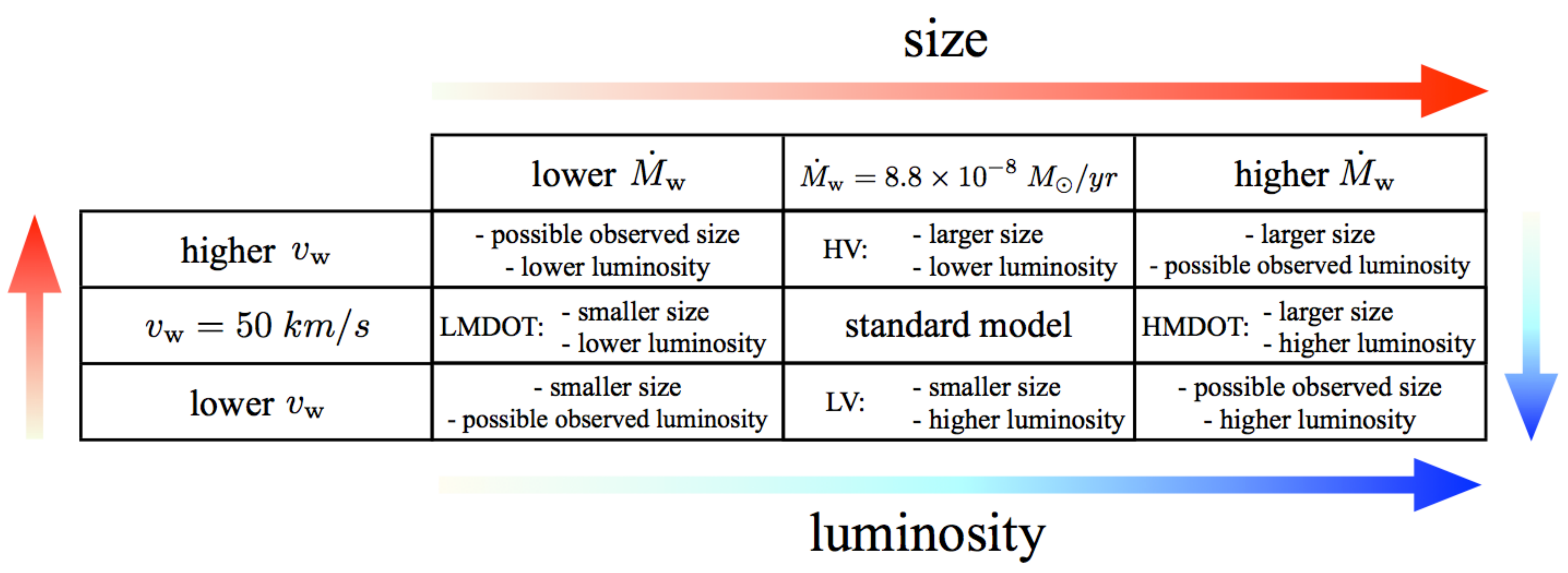}
\caption{Dependence of G2 luminosity and size on the wind parameters.
}\label{pardep}
\end{center}
\end{figure}

Our simulations show that the presence of a surrounding high-temperature atmosphere (like that predicted by ADAF/RIAF solutions for the diffuse X-ray emission in the Galactic Center) could be very important when modeling any compact source scenario for G2. As already shown by \citet{Scoville_13}, the free-streaming wind interacting with this hot atmosphere will be shocked. In the case of the so-called diffuse cloud scenario \citep{Burkert_12,Schartmann_12}, the orbital evolution of the object before pericenter is as a first approximation ballistic. In the compact source scenario, instead, due to the high thermal pressure of the ambient medium confining the outflow, the size of the free streaming wind region is always small and constrained by the equilibrium between the external pressure and the wind ram pressure. Already at early stages, a very thin, dense, and Rayleigh-Taylor unstable shell of shocked wind material forms around the free-wind region. 
The structure of the studied winds is very different from that of typical stellar winds described by \citet{Weaver_77}, where the shocked wind material forms a large shell with low density and a thinner, but dense, shell of swept and shocked ambient material propagates outwards.  In the case of the winds considered in this paper, a very weak bow shock is expected to form when the source of the wind reaches orbital velocities higher than the sound speed of the hot environment, i.e., in the late phases. Due to our numerical setup, this shell is not reproduced, but its contribution to the Br$\gamma$ luminosity of G2 is negligible, as shown by \citet{Scoville_13}. Another interesting property of this scenario is that a dominant contribution to the total luminosity comes from the shocked wind material, which has a highly filamentary structure. The shocked wind shell is in fact strongly Rayleigh-Taylor unstable due to the wind expansion and it is hence forming elongated fingers. This fact is, along with the $1/r^2$ density distribution of the free-wind region, the main difference with respect to the diffuse cloud scenario, where the object has instead a more or less uniform density all over its volume. Distinct from the diffuse cloud scenario, at late phases the ram pressure of the atmosphere can have an important role in affecting the structure of the wind (via stripping of wind material), while, as in the diffuse cloud scenario, the dominant process at late phases is the squeezing and compression by the SMBH extreme tidal field of the object in the direction parallel to the motion. A simple decoupling of all these different effects is hard to perform in an analytical study.
A slight variation of $\dot{M}\mathrm{_w}$ and $v\mathrm{_w}$ can quite significantly change the observed properties of the object. Roughly speaking, when fixing the mass-loss rates, a higher velocity results in a lower luminosity and a larger size of the emitting material (and vice versa). At constant velocity, a higher mass-loss rate instead leads to a higher luminosity and a larger size (and vice versa). Thus, a combination of observed size and luminosity can effectively constrain the wind parameters. The dependence of the luminosity and the size of the object on the wind parameters is also summarized in Figure \ref{pardep}.

Given the importance of the interaction with the hot atmosphere, we must point out that the results are of course also dependent on the model properties. For our choice \citep[which is the commonly used one; see][]{Gillessen_12, Burkert_12, Schartmann_12, Anninos_12,Scoville_13}, we found a best model, with $\dot{M}\mathrm{_w=8.8\times 10^{-8} M_{\odot} \;yr^{-1}}$ and $v\mathrm{_w = 50 \;km/s}$. These values are comparable with those of a young T Tauri star wind \citep{White_04}. The age of T Tauri stars is also consistent with the age \citep[$\simeq\mathrm{6\pm 2 \;Myr,}$][]{Paumard_06, Bartko_09} of the clockwise disk of young stars ranging from $\mathrm{0.04 \; pc}$ to $\mathrm{0.5 \; pc}$ from SgrA*, where the source is predicted to be scattered from \citep{Murray-Clay_12}.
Unfortunately, a problem for our estimates is that the Br$\gamma$ luminosity is increasing with time, particularly when the object approaches the SMBH, while the observed luminosity stayed constant from 2004 to 2012. For our best model, the corresponding luminosities range from a minimum of $\approx L_{Br\gamma,G2}/3$ to a maximum of $\approx 1.2\times L_{Br\gamma,G2}$.

\acknowledgments

This project was supported by the Deutsche Forschungsgemeinschaft priority program 1573 (“Physics of the Interstellar Medium”). Computer resources for this project have been provided by the Leibniz Supercomputing Center under grants: h0075, pr86re.

\bibliography{mylit}

\begin{thebibliography}{}
\expandafter\ifx\csname natexlab\endcsname\relax\def\natexlab#1{#1}\fi

\bibitem[{{Alig} {et~al.}(2011){Alig}, {Burkert}, {Johansson}, \&
  {Schartmann}}]{Alig_11}
{Alig}, C., {Burkert}, A., {Johansson}, P.~H., \& {Schartmann}, M. 2011,
  \mnras, 412, 469

\bibitem[{{Alig} {et~al.}(2013){Alig}, {Schartmann}, {Burkert}, \&
  {Dolag}}]{Alig_13}
{Alig}, C., {Schartmann}, M., {Burkert}, A., \& {Dolag}, K. 2013, \apj, 771,
  119

\bibitem[{{Anninos} {et~al.}(2012){Anninos}, {Fragile}, {Wilson}, \&
  {Murray}}]{Anninos_12}
{Anninos}, P., {Fragile}, P.~C., {Wilson}, J., \& {Murray}, S.~D. 2012, \apj,
  759, 132

\bibitem[{{Baganoff} {et~al.}(2003){Baganoff}, {Maeda}, {Morris}, {Bautz},
  {Brandt}, {Cui}, {Doty}, {Feigelson}, {Garmire}, {Pravdo}, {Ricker}, \&
  {Townsley}}]{Baganoff_03}
{Baganoff}, F.~K., {Maeda}, Y., {Morris}, M., {et~al.} 2003, \apj, 591, 891

\bibitem[{{Bartko} {et~al.}(2009){Bartko}, {Martins}, {Fritz}, {Genzel},
  {Levin}, {Perets}, {Paumard}, {Nayakshin}, {Gerhard}, {Alexander},
  {Dodds-Eden}, {Eisenhauer}, {Gillessen}, {Mascetti}, {Ott}, {Perrin},
  {Pfuhl}, {Reid}, {Rouan}, {Sternberg}, \& {Trippe}}]{Bartko_09}
{Bartko}, H., {Martins}, F., {Fritz}, T.~K., {et~al.} 2009, \apj, 697, 1741

\bibitem[{{Bonnet} {et~al.}(2004){Bonnet}, {Conzelmann}, {Delabre},
  {Donaldson}, {Fedrigo}, {Hubin}, {Kissler-Patig}, {Lizon}, {Paufique},
  {Rossi}, {Stroebele}, \& {Tordo}}]{Bonnet_04}
{Bonnet}, H., {Conzelmann}, R., {Delabre}, B., {et~al.} 2004, in Society of
  Photo-Optical Instrumentation Engineers (SPIE) Conference Series, ed.
  D.~{Bonaccini Calia}, B.~L. {Ellerbroek}, \& R.~{Ragazzoni}, Vol. 5490,
  130--138

\bibitem[{{Bower} {et~al.}(2003){Bower}, {Wright}, {Falcke}, \&
  {Backer}}]{Bower_03}
{Bower}, G.~C., {Wright}, M.~C.~H., {Falcke}, H., \& {Backer}, D.~C. 2003,
  \apj, 588, 331

\bibitem[{{Burkert} {et~al.}(2012){Burkert}, {Schartmann}, {Alig}, {Gillessen},
  {Genzel}, {Fritz}, \& {Eisenhauer}}]{Burkert_12}
{Burkert}, A., {Schartmann}, M., {Alig}, C., {et~al.} 2012, \apj, 750, 58

\bibitem[{{Cox} {et~al.}(2012){Cox}, {Kerschbaum}, {van Marle}, {Decin},
  {Ladjal}, {Mayer}, {Groenewegen}, {van Eck}, {Royer}, {Ottensamer}, {Ueta},
  {Jorissen}, {Mecina}, {Meliani}, {Luntzer}, {Blommaert}, {Posch},
  {Vandenbussche}, \& {Waelkens}}]{Cox_12}
{Cox}, N.~L.~J., {Kerschbaum}, F., {van Marle}, A.~J., {et~al.} 2012, \aap,
  543, C1

\bibitem[{{Cuadra} {et~al.}(2005){Cuadra}, {Nayakshin}, {Springel}, \& {Di
  Matteo}}]{Cuadra_05}
{Cuadra}, J., {Nayakshin}, S., {Springel}, V., \& {Di Matteo}, T. 2005, \mnras,
  360, L55

\bibitem[{{Cuadra} {et~al.}(2006){Cuadra}, {Nayakshin}, {Springel}, \& {Di
  Matteo}}]{Cuadra_06}
---. 2006, \mnras, 366, 358

\bibitem[{{D'Antona} \& {Mazzitelli}(1994)}]{D'Antona_94}
{D'Antona}, F., \& {Mazzitelli}, I. 1994, \apjs, 90, 467

\bibitem[{{Eisenhauer} {et~al.}(2003){Eisenhauer}, {Abuter}, {Bickert},
  {Biancat-Marchet}, {Bonnet}, {Brynnel}, {Conzelmann}, {Delabre}, {Donaldson},
  {Farinato}, {Fedrigo}, {Genzel}, {Hubin}, {Iserlohe}, {Kasper},
  {Kissler-Patig}, {Monnet}, {Roehrle}, {Schreiber}, {Stroebele}, {Tecza},
  {Thatte}, \& {Weisz}}]{Eisenhauer_03}
{Eisenhauer}, F., {Abuter}, R., {Bickert}, K., {et~al.} 2003, in Society of
  Photo-Optical Instrumentation Engineers (SPIE) Conference Series, ed.
  M.~{Iye} \& A.~F.~M. {Moorwood}, Vol. 4841, 1548--1561

\bibitem[{{Genzel} {et~al.}(2003){Genzel}, {Sch{\"o}del}, {Ott}, {Eisenhauer},
  {Hofmann}, {Lehnert}, {Eckart}, {Alexander}, {Sternberg}, {Lenzen},
  {Cl{\'e}net}, {Lacombe}, {Rouan}, {Renzini}, \& {Tacconi-Garman}}]{Genzel_03}
{Genzel}, R., {Sch{\"o}del}, R., {Ott}, T., {et~al.} 2003, \apj, 594, 812

\bibitem[{{Ghez} {et~al.}(2008){Ghez}, {Salim}, {Weinberg}, {Lu}, {Do}, {Dunn},
  {Matthews}, {Morris}, {Yelda}, {Becklin}, {Kremenek}, {Milosavljevic}, \&
  {Naiman}}]{Ghez_08}
{Ghez}, A.~M., {Salim}, S., {Weinberg}, N.~N., {et~al.} 2008, \apj, 689, 1044

\bibitem[{{Gillessen} {et~al.}(2009){Gillessen}, {Eisenhauer}, {Trippe},
  {Alexander}, {Genzel}, {Martins}, \& {Ott}}]{Gillessen_09}
{Gillessen}, S., {Eisenhauer}, F., {Trippe}, S., {et~al.} 2009, \apj, 692, 1075

\bibitem[{{Gillessen} {et~al.}(2012){Gillessen}, {Genzel}, {Fritz}, {Quataert},
  {Alig}, {Burkert}, {Cuadra}, {Eisenhauer}, {Pfuhl}, {Dodds-Eden}, {Gammie},
  \& {Ott}}]{Gillessen_12}
{Gillessen}, S., {Genzel}, R., {Fritz}, T.~K., {et~al.} 2012, \nat, 481, 51

\bibitem[{{Gillessen} {et~al.}(2013{\natexlab{a}}){Gillessen}, {Genzel},
  {Fritz}, {Eisenhauer}, {Pfuhl}, {Ott}, {Cuadra}, {Schartmann}, \&
  {Burkert}}]{Gillessen_13}
---. 2013{\natexlab{a}}, \apj, 763, 78

\bibitem[{{Gillessen} {et~al.}(2013{\natexlab{b}}){Gillessen}, {Genzel},
  {Fritz}, {Eisenhauer}, {Pfuhl}, {Ott}, {Schartmann}, {Ballone}, \&
  {Burkert}}]{Gillessen_13b}
---. 2013{\natexlab{b}}, \apj, 774, 44

\bibitem[{{Hartigan} {et~al.}(1995){Hartigan}, {Edwards}, \&
  {Ghandour}}]{Hartigan_95}
{Hartigan}, P., {Edwards}, S., \& {Ghandour}, L. 1995, \apj, 452, 736

\bibitem[{{Koo} \& {McKee}(1992)}]{Koo_92}
{Koo}, B.-C., \& {McKee}, C.~F. 1992, \apj, 388, 103

\bibitem[{{Kwak} {et~al.}(2011){Kwak}, {Henley}, \& {Shelton}}]{Kwak_11}
{Kwak}, K., {Henley}, D.~B., \& {Shelton}, R.~L. 2011, \apj, 739, 30

\bibitem[{{Lenzen} {et~al.}(1998){Lenzen}, {Hofmann}, {Bizenberger}, \&
  {Tusche}}]{Lenzen_98}
{Lenzen}, R., {Hofmann}, R., {Bizenberger}, P., \& {Tusche}, A. 1998, in
  Society of Photo-Optical Instrumentation Engineers (SPIE) Conference Series,
  ed. A.~M. {Fowler}, Vol. 3354, 606--614

\bibitem[{{Meyer} \& {Meyer-Hofmeister}(2012)}]{Meyer_12}
{Meyer}, F., \& {Meyer-Hofmeister}, E. 2012, \aap, 546, L2

\bibitem[{{Mignone} {et~al.}(2007){Mignone}, {Bodo}, {Massaglia}, {Matsakos},
  {Tesileanu}, {Zanni}, \& {Ferrari}}]{Mignone_07}
{Mignone}, A., {Bodo}, G., {Massaglia}, S., {et~al.} 2007, \apjs, 170, 228

\bibitem[{{Mignone} {et~al.}(2012){Mignone}, {Zanni}, {Tzeferacos}, {van
  Straalen}, {Colella}, \& {Bodo}}]{Mignone_12}
{Mignone}, A., {Zanni}, C., {Tzeferacos}, P., {et~al.} 2012, \apjs, 198, 7

\bibitem[{{Miralda-Escud{\'e}}(2012)}]{Miralda-Escude_12}
{Miralda-Escud{\'e}}, J. 2012, \apj, 756, 86

\bibitem[{{Murray-Clay} \& {Loeb}(2012)}]{Murray-Clay_12}
{Murray-Clay}, R.~A., \& {Loeb}, A. 2012, Nature Communications, 3,
  doi:10.1038/ncomms2044

\bibitem[{{Osterbrock} \& {Ferland}(2006)}]{Osterbrock_06}
{Osterbrock}, D.~E., \& {Ferland}, G.~J. 2006, {Astrophysics of gaseous nebulae
  and active galactic nuclei}

\bibitem[{{Parker}(1963)}]{Parker_63}
{Parker}, E.~N. 1963, {Interplanetary dynamical processes.}

\bibitem[{{Paumard} {et~al.}(2006){Paumard}, {Genzel}, {Martins}, {Nayakshin},
  {Beloborodov}, {Levin}, {Trippe}, {Eisenhauer}, {Ott}, {Gillessen}, {Abuter},
  {Cuadra}, {Alexander}, \& {Sternberg}}]{Paumard_06}
{Paumard}, T., {Genzel}, R., {Martins}, F., {et~al.} 2006, \apj, 643, 1011

\bibitem[{{Phifer} {et~al.}(2013){Phifer}, {Do}, {Meyer}, {Ghez}, {Witzel},
  {Yelda}, {Boehle}, {Lu}, {Morris}, {Becklin}, \& {Matthews}}]{Phifer_13}
{Phifer}, K., {Do}, T., {Meyer}, L., {et~al.} 2013, \apjl, 773, L13

\bibitem[{{Rousset} {et~al.}(1998){Rousset}, {Lacombe}, {Puget}, {Hubin},
  {Gendron}, {Conan}, {Kern}, {Madec}, {Rabaud}, {Mouillet}, {Lagrange}, \&
  {Rigaut}}]{Rousset_98}
{Rousset}, G., {Lacombe}, F., {Puget}, P., {et~al.} 1998, in Society of
  Photo-Optical Instrumentation Engineers (SPIE) Conference Series, ed.
  D.~{Bonaccini} \& R.~K. {Tyson}, Vol. 3353, 508--516

\bibitem[{{Schartmann} {et~al.}(2012){Schartmann}, {Burkert}, {Alig},
  {Gillessen}, {Genzel}, {Eisenhauer}, \& {Fritz}}]{Schartmann_12}
{Schartmann}, M., {Burkert}, A., {Alig}, C., {et~al.} 2012, \apj, 755, 155

\bibitem[{{Scoville} \& {Burkert}(2013)}]{Scoville_13}
{Scoville}, N., \& {Burkert}, A. 2013, \apj, 768, 108

\bibitem[{{van Marle} {et~al.}(2006){van Marle}, {Langer}, {Achterberg}, \&
  {Garc{\'{\i}}a-Segura}}]{vanMarle_06}
{van Marle}, A.~J., {Langer}, N., {Achterberg}, A., \& {Garc{\'{\i}}a-Segura},
  G. 2006, \aap, 460, 105

\bibitem[{{Vieser} \& {Hensler}(2007)}]{Vieser_07}
{Vieser}, W., \& {Hensler}, G. 2007, \aap, 472, 141

\bibitem[{{Weaver} {et~al.}(1977){Weaver}, {McCray}, {Castor}, {Shapiro}, \&
  {Moore}}]{Weaver_77}
{Weaver}, R., {McCray}, R., {Castor}, J., {Shapiro}, P., \& {Moore}, R. 1977,
  \apj, 218, 377

\bibitem[{{White} \& {Hillenbrand}(2004)}]{White_04}
{White}, R.~J., \& {Hillenbrand}, L.~A. 2004, \apj, 616, 998

\bibitem[{{Xu} {et~al.}(2006){Xu}, {Narayan}, {Quataert}, {Yuan}, \&
  {Baganoff}}]{Xu_06}
{Xu}, Y.-D., {Narayan}, R., {Quataert}, E., {Yuan}, F., \& {Baganoff}, F.~K.
  2006, \apj, 640, 319

\bibitem[{{Yuan} {et~al.}(2003){Yuan}, {Quataert}, \& {Narayan}}]{Yuan_03}
{Yuan}, F., {Quataert}, E., \& {Narayan}, R. 2003, \apj, 598, 301

\end{thebibliography}
\bibliographystyle{apj}

\end{document}